\documentclass[preprint,prd,aps,showpacs,showkeys,nofootinbib]{revtex4}
\usepackage{graphicx}
\usepackage{dcolumn}
\usepackage{bm}
\usepackage{color}
\usepackage[dvipsnames]{xcolor}
\usepackage{amssymb}
\usepackage{epstopdf}

\definecolor{black-blue}{RGB}{77,116,175}
\definecolor{black-yellow}{RGB}{231,162,33}
\definecolor{black-green}{RGB}{144,180,58}
\definecolor{black-red}{RGB}{246,95,50}

\begin{document}

\title{One-loop corrections to the neutrino in the N-B-LSSM}
\author{Shuang Di$^{1,2,3}$, Shu-Min Zhao$^{1,2,3}$\footnote{zhaosm@hbu.edu.cn}, Ming
-Yue Liu$^{1,2,3}$, Xing-Yu Han$^{1,2,3}$, Song Gao$^{1,2,3}$, Xing-Xing Dong$^{1,2,3,4}$\footnote{dongxx@hbu.edu.cn}}

\affiliation{$^1$ Department of Physics, Hebei University, Baoding 071002, China}
\affiliation{$^2$ Hebei Key Laboratory of High-precision Computation and Application of Quantum Field Theory, Baoding, 071002, China}
\affiliation{$^3$ Hebei Research Center of the Basic Discipline for Computational Physics, Baoding, 071002, China}
\affiliation{$^4$ Departamento de Fisica and CFTP, Instituto Superior T$\acute{e}$cnico, Universidade de Lisboa,
Av.Rovisco Pais 1,1049-001 Lisboa, Portugal}
\date{\today}

\begin{abstract}
In this paper, we study one-loop corrections to the neutrino mass matrix in the N-B-LSSM. We obtain the N-B-LSSM from the $U(1)$ extension of the minimal supersymmetric standard model(MSSM). By adding three generation right-handed neutrino superfields and three Higgs singlets, the model generates tiny neutrino masses at the tree level through the first type seesaw mechanism. However, one-loop corrections are essential for understanding neutrino masses and mixing angles. In our calculations, the one-loop corrections contribute approximately 10\% to the total result. We calculate the neutrino mass variance and mixing angle from both normal order neutrino mass spectrum and inverse order neutrino mass spectrum. Crucially, these corrections are 3-5 times larger than current experimental uncertainties. And they have implications for future neutrino oscillation experiments. This study provides new theoretical support for exploring the mechanism of neutrino mass generation in the supersymmetric model and provides clues for exploring new physics beyond the Standard Model (SM).
\end{abstract}

\maketitle

\section{Introduction}
The Standard Model (SM) of particle physics, despite its tremendous successes, there remains phenomena that cannot be explained. For instance, the SM fails to account for neutrino masses and their mixing patterns\cite{1sm2,2s5,3s6,4s7}. The masses and mixing patterns of neutrinos constitute a critical research area in particle physics, as they hold profound implications for cosmology and the fundamental understanding of particle interactions.

Recent experimental data, such as neutrino oscillation experiments released by the Daya Bay Laboratory in 2024\cite{DayaBay1}, have further promoted studies on neutrino masses and lepton flavor violating processes\cite{DayaBay2,DayaBay3,lep1}. These data not only provide precise measurements of neutrino mixing angles but also impose new constraints on the upper and lower bounds of neutrino masses. Such constraints are vital for unraveling the nature of neutrinos and their role in the universe. The present $3\sigma$ limits for the neutrino expriment data are\cite{6pdg}

\hspace{4cm}$0.0198 \leq \sin^{2} \theta_{13} \leq 0.024$,

\hspace{4cm}$0.268 \leq \sin^{2} \theta_{12} \leq 0.346$,

\hspace{4cm}$0.385 \leq \sin^{2} \theta_{23} \leq 0.644$,

\hspace{4cm}$6.99\times10^{-5} \leq \bigtriangleup{m^{2}}_\odot  \leq 8.07\times10^{-5}$,

\hspace{4cm}$2.371\times10^{-3} \leq \bigtriangleup{m^{2}}_A($NO$)  \leq 2.539\times10^{-3}$,

\hspace{4cm}$-2.679\times10^{-3} \leq \bigtriangleup{m^{2}}_A($IO$)  \leq -2.505\times10^{-3}$.

In the future, the precision of neutrino oscillation measurements are expected to achieve significant improvements.  Regarding future measurement sensitivities, these experiments are expected to reach sub-percent level precision for atmospheric and solar leptonic mixing parameters. This improvement will provide more rigorous tests of theoretical models that describe neutrino mass generation and mixing.
In our calculations, the one-loop corrections account for approximately 10\% to the total result. Therefore, the comprehensive one-loop corrections computed in this study are essential not only for interpreting current data but also for matching the precision requirements of next-generation experiments.

Based on the MSSM, the minimal supersymmetric extension of the SM with local B-L gauge symmetry (N-B-LSSM) is obtained by extending the gauge symmetry group to $SU(3)_C\times SU(2)_L\times U(1)_Y\times U(1)_{B-L}$\cite{7h}. The N-B-LSSM introduces three additional Higgs singlets and three generation right-handed neutrinos on the basis of the MSSM. It generates the tiny mass of light neutrinos through the seesaw mechanism. And it provides an effective solution to the $\mu$ problem that the MSSM cannot resolve. In the N-B-LSSM, the Higgs singlet $S$ acquires a non-zero vacuum expectation value (VEV) $\frac{v_S}{\sqrt{2}}$, and its superpotential $\lambda SH_uH_d$ induces an effective $\mu$ term $\mu=\frac{\lambda v_S}{\sqrt{2}}$. This mechanism replaces the $\mu H_uH_d$ term in the MSSM and alleviates the $\mu$ problem. With the introduction of three Higgs singlets, the neutral CP-even Higgs mass matrix is extended to a $5\times5$ dimension, providing a more natural explanatory framework for the 125 GeV Higgs mass.
This work assumes R-parity conservation. In the N-B-LSSM, R-parity is automatically conserved (without the need for manual addition), as the conservation law is defined as: $R_p=(-1)^{3(B-L)+2S}$.
Other terms in superpotential are allowed. For convenience, they are not considered in this study, because their effects can be combined in the already present terms, and they do not have a substantial new impact on our research.

In the N-B-LSSM, we have discussed certain two-loop contributions to muon anomalous magnetic dipole moment, lepton flavour-violating decays $l_j\rightarrow l_{i\gamma},l_j\rightarrow 3l_i$ and $\mu\rightarrow e+q\overline{q}$, but the neutrino mass problem has not yet been studied\cite{7h}. In this paper, we study the neutrino mass correction problem under the N-B-LSSM, neutrinos acquire tiny masses at tree level through the seesaw mechanism\cite{8seesaw,9sm7,102zlxzqqb,11Minkowski}, facilitated by the introduction of right-handed neutrino fields. However, the one-loop corrections are crucial not only for neutrino mass and mixing\cite{10yyl22,11sm} but also significantly impact other physical processes. Loop corrections play essential roles in lepton flavor violating processes such as the radiative decays $l_j\rightarrow l_i\gamma~(i,~j=1,~2,~3,~i\neq j)$\cite{lep1}
  with zero tree-level contribution. Neutrinos have tiny masses, which implies lepton flavor violation.
  Therefore, we take into account the important one-loop corrections in this work. Our corrections arise from various sources, including virtual slepton-chargino, sneutrino-neutralino, and Higgs-charged lepton loops. The inclusion of one-loop effects is essential to match the precise experimental data on neutrino oscillations, which provide stringent constraints on the neutrino mass squared differences and mixing angles.

In Ref.\cite{12tff}, the authors discuss the supersymmetric neutrino mass under the one-loop approximation using the on-shell renormalization scheme without taking into account lepton number conservation and R-parity. There are also some work related to the mass and mixing of neutrinos\cite{13yyl,101zlxz,14yyl33,15tff8}. This paper focuses on the one-loop corrections to the neutrino mass matrix in the N-B-LSSM. We employ the method of mass eigenstates to calculate these corrections and analyze their impact on the neutrino mass spectrum and mixing angles. Our study aims to explore the parameter space of the N-B-LSSM that is consistent with the current neutrino oscillation data. By doing so, we aim to shed light on the underlying mechanisms of neutrino mass generation and to provide insights into the new physics beyond the SM.

The structure of this paper is as follows: In Sec.II, we briefly review the N-B-LSSM and its key features relevant to neutrino mass generation. Sec.III details the calculation of one-loop corrections to the neutrino mass matrix. In Sec.IV, we present our numerical results and discuss the implications for the neutrino mass spectrum and mixing angles. Finally, the conclusion are given out in Sec.V.

\section{The N-B-LSSM}
N-B-LSSM is the $U(1)_{B-L}$  extension of MSSM, and the local gauge group is {$SU(3)_C\times SU(2)_L \times U(1)_Y\times U(1)_{B-L}$}\cite{16s13}. In order to obtain the N-B-LSSM, new superfields are added to the MSSM, including three Higgs singlets $\hat{\chi}_1,~\hat{\chi}_2,~\hat{S}$ and right-handed superfields $\hat{\nu}_i$. Through the seesaw mechanism, the minuteness of the neutrino mass can be naturally explained. The neutral CP-even scalar fields of  $H_u$, $H_d$, $\chi_1$,$~\chi_2$ and $S$ mix into a $5\times5$ mass squared matrix.
The superpotential in N-B-LSSM is expressed as
\begin{eqnarray}
&&W=-Y_d\hat{d}\hat{q}\hat{H}_d-Y_e\hat{e}\hat{l}\hat{H}_d-\lambda_2\hat{S}\hat{\chi}_1\hat{\chi}_2+\lambda\hat{S}\hat{H}_u\hat{H}_d
\nonumber\\&&~~~~~~~+\frac{\kappa}{3}\hat{S}\hat{S}\hat{S}+Y_u\hat{u}\hat{q}\hat{H}_u+Y_{X}\hat{\nu}\hat{\chi}_1\hat{\nu}+Y_\nu\hat{\nu}\hat{l}\hat{H}_u.
\end{eqnarray}

Under this model, $Y_{u,d,e,\nu,\chi}$ in the superpotential denote the Yukawa coupling coefficients. While $\lambda$, $\lambda_2$ and $\kappa$ are dimensionless couplings constant. $\hat{\chi}_1,~\hat{\chi}_2,~\hat{S}$ are three Higgs singlets.

The vacuum expectation values(VEVs) of the Higgs superfields $H_u$, $H_d$, $\chi_1$, $\chi_2$ and $S$ are presented by $v_u,~v_d,~v_\eta$,~ $v_{\bar\eta}$ and $v_S$ respectively. Two angles are defined as $\tan\beta=v_u/v_d$ and $\tan\beta_\eta=v_{\bar{\eta}}/v_{\eta}$. The definition of $\tilde{\nu}_L$ and $\tilde{\nu}_R$ is
\begin{eqnarray}
&&\hspace{-2cm}\tilde{\nu}_L={{1\over\sqrt{2}}\phi_L+{i\over\sqrt{2}}\sigma_L},   ~~~~~~
\tilde{\nu}_R={{1\over\sqrt{2}}\phi_R+{i\over\sqrt{2}}\sigma_R}.
\end{eqnarray}

The specific explicit expressions of the neutral
components of the two Higgs doublets and three Higgs singlets are as follows
\begin{eqnarray}
&&\hspace{-2cm}H^0_d={1\over\sqrt{2}}\phi_{d}+{1\over\sqrt{2}}v_{d}+i{1\over\sqrt{2}}\sigma_d,
\nonumber\\&&\hspace{-2cm}H^0_u={1\over\sqrt{2}}\phi_{u}+{1\over\sqrt{2}}v_{u}+i{1\over\sqrt{2}}\sigma_u,
\nonumber\\&&\hspace{-2cm}\chi_1={1\over\sqrt{2}}\phi_{1}+{1\over\sqrt{2}}v_{\eta}+i{1\over\sqrt{2}}\sigma_1,
\nonumber\\&&\hspace{-2cm}\chi_2={1\over\sqrt{2}}\phi_{2}+{1\over\sqrt{2}}v_{\bar{\eta}}+i{1\over\sqrt{2}}\sigma_2,
%\nonumber\\&&\hspace{-2cm}\tilde{\nu}_L={1\over\sqrt{2}}\phi_L+i{1\over\sqrt{2}}\sigma_L,
%\nonumber\\&&\hspace{-2cm}\tilde{\nu}_R={1\over\sqrt{2}}\phi_R+i{1\over\sqrt{2}}\sigma_R,
\nonumber\\&&\hspace{-2cm}S={1\over\sqrt{2}}\phi_S+{1\over\sqrt{2}}v_S+i{1\over\sqrt{2}}\sigma_S.
\end{eqnarray}

The soft SUSY breaking terms are shown as
\begin{eqnarray}
&&\mathcal{L}_{soft}=\mathcal{L}_{soft}^{MSSM}-\frac{T_\kappa}{3}S^3+\epsilon_{ij}T_{\lambda}SH_d^iH_u^j+T_{2}S\chi_1\chi_2\nonumber\\&&
-T_{\chi,ik}\chi_1\tilde{\nu}_{R,i}^{*}\tilde{\nu}_{R,k}^{*}
+\epsilon_{ij}T_{\nu,ij}H_u^i\tilde{\nu}_{R,i}^{*}\tilde{e}_{L,j}-m_{\eta}^2|\chi_1|^2-m_{\bar{\eta}}^2|\chi_2|^2\nonumber\\&&-m_S^2|S|^2-m_{\nu,ij}^2\tilde{\nu}_{R,i}^{*}\tilde{\nu}_{R,j}
-\frac{1}{2}(2M_{BB^\prime}\lambda_{\tilde{B}}\tilde{B^\prime}+\delta_{ij} M_{BL}\tilde{B^\prime}^2)+h.c~~,
\end{eqnarray}
$\mathcal{L}_{soft}^{MSSM}$ represent the soft breaking terms of the MSSM. The parameters $T_{\kappa}$, $T_{\lambda}$, $T_2$, $T_{\chi}$ and $T_{\nu}$ are trilinear coupling coefficients.
\begin{table}[h]
\caption{ The superfields in N-B-LSSM}
\begin{tabular}{|c|c|c|c|c|}
\hline
Superfields & $U(1)_Y$ & $SU(2)_L$ & $SU(3)_C$ & $U(1)_{B-L}$ \\
\hline
$\hat{q}$ & 1/6 & 2 & 3 & 1/6  \\
\hline
$\hat{l}$ & -1/2 & 2 & 1 & -1/2  \\
\hline
$\hat{H}_d$ & -1/2 & 2 & 1 & 0 \\
\hline
$\hat{H}_u$ & 1/2 & 2 & 1 & 0 \\
\hline
$\hat{d}$ & 1/3 & 1 & $\bar{3}$ & -1/6  \\
\hline
$\hat{u}$ & -2/3 & 1 & $\bar{3}$ & -1/6 \\
\hline
$\hat{e}$ & 1 & 1 & 1 & $1/2$  \\
\hline
$\hat{\nu}$ & 0 & 1 & 1 & $1/2$ \\
\hline
$\hat{\chi}_1$ & 0 & 1 & 1 & -1 \\
\hline
$\hat{\chi}_2$ & 0 & 1 & 1 & 1\\
\hline
$\hat{S}$ & 0 & 1 & 1 & 0 \\
\hline
\end{tabular}
\label{table1}
\end{table}
The particle content and charge assignments of N-B-LSSM are shown in the Table  \ref {table1}. In chiral superfield frameworks, $\hat H_u = \Big( {\hat H_u^ + ,\hat H_u^0} \Big)$ and $\hat H_d = \Big( {\hat H_d^0,\hat H_d^ - } \Big)$ represent the MSSM-like doublet Higgs superfields. $\hat q $ and $\hat l $ are the doublets of quark and lepton. $\hat u$, $\hat d$, $\hat e$ and $\hat{\nu}$ are the singlet up-type quark, down-type quark, charged lepton and neutrino superfields, respectively.

The gauge groups $U(1)_Y$ and {$U(1)_{B-L}$} exhibit a gauge kinetic mixing effect, which can also be induced through renormalization group equations(RGEs)even with a zero value at $M_{GUT}$. Because the two Abelian gauge groups remain unbroken, a basis conversion can occur using a rotation matrix $R$ ($R^T R=1$)\cite{17h32,18h33,19h34,20h35}. The covariant derivatives of this model can be expressed as follows
 {\begin{eqnarray}
%%%%%%%%%%%%%%%%%%%%%%%%%%%%%%%%%%%%%%%%%%%%%%%%%%%%%%%%%%%%%%%%%%%%
&&D_\mu=\partial_\mu-i\left(\begin{array}{cc}Y,&B-L\end{array}\right)
\left(\begin{array}{cc}g_{Y},&g{'}_{{YB}}\\g{'}_{{BY}},&g{'}_{{B-L}}\end{array}\right)
\left(\begin{array}{c}B_{\mu}^{\prime Y} \\ B_{\mu}^{\prime BL}\end{array}\right)\;,
%%%%%%%%%%%%%%%%%%%%%%%%%%%%%%%%%%%%%%%%%%%%%%%%%%%%%%%%%%%%%%%%%%%%
\label{gauge1}
\end{eqnarray}}
in which $Y$ and $B-L$ denote the hypercharge and the $B-L$ (baryon number minus lepton number)charge, respectively. $g_B$ represents the gauge coupling constant of the {$U(1)_{B-L}$} group. $g_{YB}$ represents the mixing gauge coupling constant of the  {$U(1)_{B-L}$} group and $U(1)_Y$ group. Given that the two Abelian gauge symmetries remain unbroken, a basis transformation can be executed
\begin{eqnarray}
%%%%%%%%%%%%%%%%%%%%%%%%%%%%%%%%%%%%%%%%%%%%%%%%%%%%%%%%%%%%%%%%%%%%
&&\left(\begin{array}{cc}g_{Y},&g{'}_{{YB}}\\g{'}_{{BY}},&g{'}_{{B-L}}\end{array}\right)
R^T=\left(\begin{array}{cc}g_{1},&g_{{YB}}\\0,&g_{{B}}\end{array}\right)\;.
%%%%%%%%%%%%%%%%%%%%%%%%%%%%%%%%%%%%%%%%%%%%%%%%%%%%%%%%%%%%%%%%%%%%
\label{gauge3}
\end{eqnarray}

Finally, the gauge derivative of N-B-LSSM is transformed into
{\begin{eqnarray}
%%%%%%%%%%%%%%%%%%%%%%%%%%%%%%%%%%%%%%%%%%%%%%%%%%%%%%%%%%%%%%%%%%%%
&&R\left(\begin{array}{c}B_{\mu}^{\prime Y} \\ B_{\mu}^{\prime BL}\end{array}\right)
=\left(\begin{array}{c}B_{\mu}^{Y} \\ B_{\mu}^{BL}\end{array}\right)\;.
%%%%%%%%%%%%%%%%%%%%%%%%%%%%%%%%%%%%%%%%%%%%%%%%%%%%%%%%%%%%%%%%%%%%
\label{gauge4}
\end{eqnarray}}

The mass matrix for neutralino in the basis $(\lambda_{\tilde{B}}, \tilde{W}^0, \tilde{H}_d^0, \tilde{H}_u^0,
\tilde{B'}, \tilde{\chi_1}, \tilde{\chi_2}, S) $ is

\begin{eqnarray}
m_{\chi^0} = \left(
\begin{array}{cccccccc}
M_1 &0 &-\frac{1}{2}g_1 v_d &\frac{1}{2} g_1 v_u &{M}_{B B'} & 0  & 0  &0\\
0 &M_2 &\frac{1}{2} g_2 v_d  &-\frac{1}{2} g_2 v_u  &0 &0 &0 &0\\
-\frac{1}{2}g_1 v_d &\frac{1}{2} g_2 v_d  &0
&- \frac{1}{\sqrt{2}} {\lambda} v_S&-\frac{1}{2} g_{YB} v_d &0 &0 & - \frac{1}{\sqrt{2}} {\lambda} v_u\\
\frac{1}{2}g_1 v_u &-\frac{1}{2} g_2 v_u  &- \frac{1}{\sqrt{2}} {\lambda} v_S &0 &\frac{1}{2} g_{YB} v_u  &0 &0 &- \frac{1}{\sqrt{2}} {\lambda} v_d\\
{M}_{B B'} &0 &-\frac{1}{2} g_{YB} v_{d}  &\frac{1}{2} g_{YB} v_{u} &{M}_{BL} &- g_{B} v_{\eta}  &g_{B} v_{\bar{\eta}}  &0\\
0  &0 &0 &0 &- g_{B} v_{\eta}  &0 &-\frac{1}{\sqrt{2}} {\lambda}_{2} v_S  &-\frac{1}{\sqrt{2}} {\lambda}_{2} v_{\bar{\eta}} \\
0 &0 &0 &0 &g_{B} v_{\bar{\eta}}  &-\frac{1}{\sqrt{2}} {\lambda}_{2} v_S  &0 &-\frac{1}{\sqrt{2}} {\lambda}_{2} v_{\eta} \\
0 &0 & - \frac{1}{\sqrt{2}} {\lambda} v_u &- \frac{1}{\sqrt{2}}{\lambda} v_d &0 &-\frac{1}{\sqrt{2}} {\lambda}_{2} v_{\bar{\eta}}
 &-\frac{1}{\sqrt{2}} {\lambda}_{2} v_{\eta}  &\sqrt{2}\kappa v_S\end{array}
\right).\label{neutralino}
 \end{eqnarray}

This matrix is diagonalized by the rotation matrix $N$
\begin{eqnarray}
N^{*}m_{\chi^0} N^{\dagger} = m^{diag}_{\chi^0}.
\end{eqnarray}

One can find other mass matrixes in the Appendix \ref{A1}.
\section{ANALYTICAL FORMULA}
In this section, we compute the one-loop correction to the neutrino mass under the N-B-LSSM. We perform the chiral decomposition of the fermion propagator in loop diagrams, where $P_L=\frac{1-\gamma_5}{2}$ and $P_R=\frac{1+\gamma_5}{2}$ present chiral projection operator. The general form of the self-energy of $\nu_{i}-\nu_{j}$ is as follows
\begin{eqnarray}
S_{ij}(k)={a_{ij}m_jP_L}+{b_{ij}m_iP_R}+{c_{ij}{k\!\!\!\slash}P_L}+{d_{ij}{k\!\!\!\slash}P_R}.
\label{e111}\end{eqnarray}

Considering the one-loop correction, $a_{ij},b_{ij},c_{ij}$ and $d_{ij}$ are functions of $k^2$. In  the one-loop correction calculation,
  $k^2$ is the external neutrino momentum square, which is small.
  So, we perform a Taylor expansion of $a_{ij},b_{ij},c_{ij}$ and $d_{ij}$ according to $k^2$. Taking into account the leading-order and next-to-leading order contributions of $a_{ij},b_{ij},c_{ij}$ and $d_{ij}$, we obtain\cite{12tff,21tff12}
\begin{eqnarray}
&&\hspace{-2cm}a_{ij}=a^0_{ij}+k^2a^1_{ij},
\nonumber\\&&\hspace{-2cm}b_{ij}=b^0_{ij}+k^2b^1_{ij},
\nonumber\\&&\hspace{-2cm}c_{ij}=c^0_{ij}+k^2c^1_{ij},
\nonumber\\&&\hspace{-2cm}d_{ij}=d^0_{ij}+k^2d^1_{ij}.
\label{e222}\end{eqnarray}

By adding counter-terms $S(k)'_{ij}$, we renormalize $S(k)_{ij}$ as
\begin{eqnarray}
S_{ij}^{REN}(k)=S(k)'_{ij}+{a_{ij}^*m_jP_L}+{b_{ij}^*m_iP_R}+
{c_{ij}^*{k\!\!\!\slash}P_L}+{d_{ij}^*{k\!\!\!\slash}P_R}
,\label{e12}
\end{eqnarray}
where the parts with * are the counter parts. In the used renormalization scheme they are determined by the mass-shell conditions
\begin{eqnarray}
&&\hspace{-2cm}S_{ij}^{REN}(k)u_j(k)|_{k^2=m_j^2}=0,
\nonumber\\&&\hspace{-2cm}\bar{u}_i(k)S_{ij}^{REN}(k)|_{k^2=m_i^2}=0.
\end{eqnarray}
From Eqs. (11)(12)(13), we can get the following solution
\begin{eqnarray}
&&\hspace{0cm}a_{ij}^*=-a_{ij}^0+{m_i}^2b_{ij}^1+{m_i}^2c_{ij}^1+{m_i}^2{m_j}^2d_{ij}^1,
\nonumber\\&&\hspace{0cm}b_{ij}^*=-b_{ij}^0+{m_i}^2a_{ij}^1+{m_i}^2c_{ij}^1+{m_i}^2{m_j}^2d_{ij}^1,
\nonumber\\&&\hspace{0cm}c_{ij}^*=-c_{ij}^0-{m_j}^2a_{ij}^1-{m_i}^2b_{ij}^1-({m_i}^2+{m_j}^2)c_{ij}^1-{m_i}^2{m_j}^2d_{ij}^1,
\nonumber\\&&\hspace{0cm}d_{ij}^*=-d_{ij}^0-{m_i}^2{m_j}^2a_{ij}^1-{m_i}^2{m_j}^2b_{ij}^1-{m_i}^2{m_j}^2c_{ij}^1-({m_i}^2+{m_j}^2)d_{ij}^1.
\label{e14}
\end{eqnarray}

According to Eq.(\ref{e12}) and Eq.(\ref{e14}), the renormalized self-energy can be written as
\begin{eqnarray}
&&\hspace{0cm}S_{ij}^{REN}(k)=({m_i}^2b_{ij}^1+{m_i}^2c_{ij}^1+{m_i}^2{m_j}^2d_{ij}^1+a_{ij}^1k^2)m_jP_L
\nonumber\\&&\hspace{1.5cm}+({m_j}^2a_{ij}^1+{m_j}^2c_{ij}^1+{m_i}^2{m_j}^2d_{ij}^1+b_{ij}^1k^2)m_iP_R
\nonumber\\&&\hspace{1.5cm}+\Big(-{m_j}^2a_{ij}^1-{m_i}^2b_{ij}-({m_i}^2+{m_j}^2)c_{ij}^1-{m_i}^2{m_j}^2d_{ij}^1+b_{ij}^1k^2\Big)k\!\!\!\slash P_L
\nonumber\\&&\hspace{1.5cm}+\Big(-{m_i}^2{m_j}^2a_{ij}^1-{m_i}^2{m_j}^2b_{ij}^1-{m_i}^2{m_j}^2c_{ij}^1-({m_i}^2+{m_j}^2)f_{ij}^1+d_{ij}^1k^2\Big)k\!\!\!\slash P_R
\nonumber\\&&\hspace{1.5cm}=(k\!\!\!\slash-m_j)\hat{S}_{ij}(k)(k\!\!\!\slash-m_i),
\end{eqnarray}
where $\hat{S}_{ij}(k)$ is denoted as
\begin{eqnarray}
&&\hspace{0cm}\hat{S}_{ij}(k)=a_{ij}^1m_jP_R+b_{ij}^1m_iP_L+c_{ij}^1(m_iP_L+m_jP_R+k\!\!\!\slash P_R)
\nonumber\\&&\hspace{1.5cm}+d_{ij}^1(m_iP_R+m_jP_R+k\!\!\!\slash P_L).
\end{eqnarray}

For convenience, we have introduced some new symbols
\begin{eqnarray}
&&\hspace{0cm}\delta{Z_{ij}^L}=-{m_j}^2a_{ij}^1-{m_i}^2b_{ij}-({m_i}^2+{m_j}^2)c_{ij}^1-{m_i}^2{m_j}^2d_{ij}^1+b_{ij}^1k^2,
\nonumber\\&&\hspace{0cm}\delta{Z_{ij}^R}=-{m_i}^2{m_j}^2a_{ij}^1-{m_i}^2{m_j}^2b_{ij}^1-{m_i}^2{m_j}^2c_{ij}^1-({m_i}^2+{m_j}^2)f_{ij}^1+d_{ij}^1k^2,
\nonumber\\&&\hspace{0cm}\delta{m_{ij}^L}=({m_i}^2b_{ij}^1+{m_i}^2c_{ij}^1+{m_i}^2{m_j}^2d_{ij}^1+a_{ij}^1k^2)m_j,
\nonumber\\&&\hspace{0cm}\delta{m_{ij}^R}=({m_j}^2a_{ij}^1+{m_j}^2c_{ij}^1+{m_i}^2{m_j}^2d_{ij}^1+b_{ij}^1k^2)m_i,
\end{eqnarray}
where $\delta{Z_{ij}^L}(\delta{Z_{ij}^R})$ is left-handed(right-handed) wave function renormalization constant.
 $\delta{m_{ij}^L}(\delta{m_{ij}^R})$ is left-handed(right-handed) mass renormalization constant. They are used to construct the renormalization propagator.

Under the one-loop order, the Green's function at two points is expressed as
\begin{eqnarray}
&&\hspace{0cm}\Gamma_{ij}(k)=(k\!\!\!\slash-m_i^{tr})\delta_{ij}+S_{ij}^{REN}(k)
\nonumber\\&&\hspace{0cm}= (\delta_{ij} + \delta Z_{ij}^L)(k\!\!\!\slash - m_i^{tr} - \delta m_{ij}^L + \delta Z_{ij}^Lm_i^{tr})P_L
\nonumber\\&&\hspace{0cm}+ (\delta_{ij} + \delta Z_{ij}^R)(k\!\!\!\slash - m_j^{tr} - \delta m_{ij}^R + \delta Z_{ij}^R m_j^{tr})P_R,
\end{eqnarray}
where $\delta_{ij} + \delta Z_{ij}^L$ stands for the left-handed wave function's renormalization multiplier and $\delta_{ij} + \delta Z_{ij}^R$ stands for the right-handed wave function's renormalization multiplier. $m_i^{tr}$  refers to the mass of the ith generation fermion at the tree level. Based on Eq.(18) and the mass-shell conditions, we obtain the loop corrections for the mass matrix elements as follows
\begin{eqnarray}
&&\hspace{0cm}\delta m^{loop}_{ij}=3m_i^{tr}(m_i^{tr})^2a_{ij}^1+\Big(m_i^{tr}m_j^{tr}+(m_i^{tr})^2+(m_j^{tr})^2\Big)m_i^{tr}b_{ij}^1
\nonumber\\&&\hspace{1.2cm}+\Big((m_i^{tr})^2m_j^{tr}+3m_i^{tr}(m_j^{tr})^2\Big)c_{ij}^1+\Big(3(m_i^{tr})^2m_j^{tr}+m_i^{tr}(m_i^{tr})^2\Big)d_{ij}^1.
\end{eqnarray}

\begin{figure}[h]
\setlength{\unitlength}{5mm}
\centering
\includegraphics[width=3.5in]{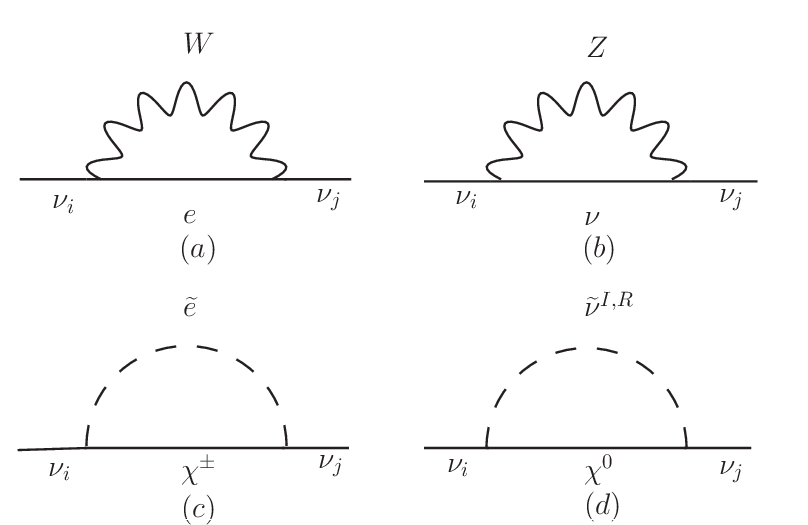}
\setlength{\unitlength}{5mm}
\caption{The one-loop self-energy diagrams}{\label {T111}}
\label{t1}
\end{figure}

To compute $a_{ij}^0,b_{ij}^0,c_{ij}^0,d_{ij}^0$ , the one-loop self energy diagrams should be precisely calculated. The associated Feynman diagrams are shown in Fig.\ref{T111}. The exchanged bosons can be of vectorial or scalar type, which leads to different integral calculations. The integral for exchanging vector-boson is
\begin{eqnarray}
&&\hspace{0cm}\mathcal{M}_\nu^{a,b}=-\bar{u}(P)\int\frac{d^Dk}{(2\pi)^D}i\gamma_\mu(A_LP_L+A_RP_R)\frac{i}{{k\!\!\!\slash}-m_F}i\gamma_\beta(B_LP_L+B_RP_R)\frac{-ig^{\alpha\beta}}{(p-k)^2-m_V^2}u(p)
\nonumber\\&&\hspace{0.8cm}=-\int^1_0dx\int\frac{d^Dk}{(2\pi)^D}\frac{1}{[k^2-xm_V^2-(1-x)m_F^2]}[1+\frac{2x(1-x)p^2}{k^2+xm_V^2+(1-x)m_F^2}]
\nonumber\\&&\hspace{0.8cm}\times[A_RB_LP_LDm_F+A_LB_RP_RDm_F+A_LB_LP_L(2-D)p\!\!\!\slash x+A_RB_RP_R(2-D)p\!\!\!\slash x],
\label{e333}
\end{eqnarray}
where $D=4-2\varepsilon$. $m_V$ denotes the mass of the vector boson appearing in the loop, and $m_F$ denotes the fermion in the loop. From Eq.(\ref{e111}), Eq.(\ref{e222}) and Eq.(\ref{e333}), we obtain
\begin{eqnarray}
&&\hspace{0cm}a_{ij}^0(m_V,m_F)=iD\frac{m_F}{m_j}A_RB_LF_1(m_F,m_V),
\nonumber\\&&\hspace{0cm}b_{ij}^0(m_V,m_F)=i\frac{m_F}{m_j}A_LB_RF_1(m_F,m_V),
\nonumber\\&&\hspace{0cm}c_{ij}^0(m_V,m_F)=iA_LB_LF_2(m_F,m_V),
\nonumber\\&&\hspace{0cm}d_{ij}^0(m_V,m_F)=iA_RB_RF_2(m_F,m_V),
\nonumber\\&&\hspace{0cm}a_{ij}^1(m_V,m_F)=i\frac{m_F}{m_j}A_RB_LF_3(m_F,m_V),
\nonumber\\&&\hspace{0cm}b_{ij}^1(m_V,m_F)=i\frac{m_F}{m_j}A_LB_RF_3(m_F,m_V),
\nonumber\\&&\hspace{0cm}c_{ij}^1(m_V,m_F)=iA_RB_LF_4(m_F,m_V),
\nonumber\\&&\hspace{0cm}d_{ij}^1(m_V,m_F)=iA_LB_RF_4(m_F,m_V),
\end{eqnarray}
where $F_1$, $F_2$, $F_3$ and $F_4$ are integrals over the internal momentum of the loop and their explicit forms are given in appendix \ref{B1}. In Fig.(\ref{t1})(b), the specific forms of $A_L^{Z,\nu}$,$A_R^{Z,\nu}$,$B_L^{Z,\nu}$ and $B_R^{Z,\nu}$ are
\begin{eqnarray}
&&\hspace{2.5cm}A_L^{Z,\nu}=-\frac{i}{2}(g_1\cos\theta'_w\sin\theta_w+g_2\cos\theta_w\cos\theta'_w)\sum^{3}_{a=1}U_{j,a}^{V,*}U_{i,a}^{V},
\nonumber\\&&\hspace{2.5cm}A_R^{Z,\nu}=\frac{i}{2}(g_1\cos\theta'_w\sin\theta_w+g_2\cos\theta_w\cos\theta'_w)\sum^{3}_{a=1}U_{i,a}^{V,*}U_{j,a}^{V},
\nonumber\\&&\hspace{2.5cm}B_L^{Z,\nu}=\frac{i}{2}\Big((g_1\sin\theta_w+g_2\cos\theta_w)\sin\theta'_w+(g_{YB}+g_B)\cos\theta'_w\Big)\sum^{3}_{a=1}U_{j,a}^{V,*}U_{i,a}^{V}
\nonumber\\&&\hspace{3cm}-g_B\cos\theta'_w\sum^{3}_{a=1}U_{j,3+a}^{V,*}U_{i,3+a}^{V},
\nonumber\\&&\hspace{2.5cm}B_R^{Z,\nu}=-\frac{i}{2}\Big((g_1\sin\theta_w+g_2\cos\theta_w)\sin\theta'_w+(g_{YB}+g_B)\cos\theta'_w\Big)\sum^{3}_{a=1}U_{i,a}^{V,*}U_{j,a}^{V}
\nonumber\\&&\hspace{3cm}-g_B\cos\theta'_w\sum^{3}_{a=1}U_{i,3+a}^{V,*}U_{j,3+a}^{V}.
\end{eqnarray}

$\theta_w$ denotes the Weinberg angle, and $\theta'_w$ denotes $B-L$ mixing angle.
$U^V$ denotes the rotation matrix to diagonalize the neutrino mass  matrix.

For the exchange of scalar bosons, the amplitude is derived in a similar manner. And it is written as
\begin{eqnarray}
&&\hspace{0cm}\mathcal{M}_\nu^{c,d}=-\bar{u}(P)\int\frac{d^Dk}{(2\pi)^D}i(C_LP_L+C_RP_R)\frac{i}{{k\!\!\!\slash}-m_F}i(D_LP_L+D_RP_R)\frac{-i}{(p-k)^2-m_S^2}u(p)
\nonumber\\&&\hspace{0.8cm}=-\int^1_0dx\int\frac{d^Dk}{(2\pi)^D}\frac{1}{[k^2-xm_S^2-(1-x)m_F^2]}[1+\frac{2x(1-x)p^2}{k^2+xm_S^2+(1-x)m_F^2}]
\nonumber\\&&\hspace{0.8cm}\times[(C_RD_LP_L+C_LD_RP_R)m_F+(C_LD_LP_L x+C_RD_RP_R)p\!\!\!\slash x],
\label{e444}
\end{eqnarray}
where $m_S$ denotes the mass of the scalar boson appearing in the loop, and $m_F$ denotes the fermion in the loop. From Eq.(\ref{e111}), Eq.(\ref{e222}) and Eq.(\ref{e444}), we obtain
\begin{eqnarray}
&&\hspace{0cm}a_{ij}^0(m_S,m_F)=i\frac{m_F}{m_j}C_LD_LF_1(m_F,m_S),
\nonumber\\&&\hspace{0cm}b_{ij}^0(m_S,m_F)=i\frac{m_F}{m_j}C_RD_RF_1(m_F,m_S),
\nonumber\\&&\hspace{0cm}c_{ij}^0(m_S,m_F)=iC_RD_LF_2(m_F,m_S),
\nonumber\\&&\hspace{0cm}d_{ij}^0(m_S,m_F)=iC_LD_RF_2(m_F,m_S),
\nonumber\\&&\hspace{0cm}a_{ij}^1(m_S,m_F)=i\frac{m_F}{m_j}C_LD_LF_3(m_F,m_S),
\nonumber\\&&\hspace{0cm}b_{ij}^1(m_S,m_F)=i\frac{m_F}{m_j}C_LD_LF_3(m_F,m_S),
\nonumber\\&&\hspace{0cm}c_{ij}^1(m_S,m_F)=iC_RD_LF_4(m_F,m_S),
\nonumber\\&&\hspace{0cm}d_{ij}^1(m_S,m_F)=iC_LD_RF_4(m_F,m_S).
\end{eqnarray}

In Fig.(\ref{t1})(c), the specific forms of $C_L^{\chi^\pm,e,\nu}$,$C_R^{\chi^\pm,e,\nu}$,$D_L^{\chi^\pm,e,\nu}$ and $D_R^{\chi^\pm,e,\nu}$ are
\begin{eqnarray}
&&\hspace{2.5cm}C_L^{\chi^\pm,e,\nu}=-g_2U_{j,1}^*\sum^{3}_{a=1}U_{i,a}^{V,*}Z_{k,a}^E+U_{j,2}\sum^{3}_{a=1}U_{i,b}^{V,*}\sum^{3}_{a=1}Y_{e,ab}Z_{k,3+a}^E,
\nonumber\\&&\hspace{2.5cm}C_R^{\chi^\pm,e,\nu}=\sum^{3}_{b=1}\sum^{3}_{a=1}Y_{\nu,ab}^*U_{i,3+a}^VZ_{k,b}^EV_{j,2},
\nonumber\\&&\hspace{2.5cm}D_L^{\chi^\pm,e,\nu}=-g_2\sum^{3}_{a=1}Z_{k,a}^{E,*}U_{j,a}^VU_{i,1}+\sum^{3}_{a=1}\sum^{3}_{b=1}Y_{e,ab}^*Z_{k,3+a}^{E,*}U_{j,b}^VU_{i,2},
\nonumber\\&&\hspace{2.5cm}D_R^{\chi^\pm,e,\nu}=V_{i,2}^*\sum^{3}_{b=1}Z_{k,b}^{E,*}\sum^{3}_{a=1}U_{j,3+a}^{V,*}Y_{\nu,ab}.
\end{eqnarray}
$Z^E$ denotes the rotation matrix to diagonalize the slepton mass squared matrix. We introduce them in detail in appendix \ref{A1}.

At tree level, the neutrino mass mixing matrix is
\begin{eqnarray}
M_\nu=&&\left(\begin{array}{cc}0&\frac{1}{\sqrt{2}}{v}_{u} Y_{\nu}^T
\\ \frac{1}{\sqrt{2}}v_{u}Y_{\nu}&\sqrt{2}{v}_{\eta}Y_X\end{array}\right)\;.
\end{eqnarray}

The masses of neutrinos are obtained through the rotation matrix $U^V$ using the formula $(U^{V})^TM_\nu U^V=diag(m_{\nu^\alpha})$, where $\alpha=1...6$. Here, $M_\nu$ is the mass matrix of neutrinos,
and $diag(m_{\nu^\alpha})$ is the diagonalized mass matrix.
 We use the matrix $(U^{V})^T$ in the leading order of $\zeta=\frac{v_u}{{v}_{\eta}}(Y_{\nu}^T)_{3\times3}\cdot(Y_X)_{3\times3}^{-1}$. Since the neutrino Yukawa coupling $Y_\nu$ is very small, all elements of $\zeta$ are much smaller than 1 ($\zeta_{IJ}\ll1,~I,~J=1,~2,~3$). In the leading-order approximation of $\zeta$, $(U^{V})^T$ can be expressed as
\begin{eqnarray}
&(U^{V})^T=\left(\begin{array}{cc}\mathcal{S}^T&0\\
0&\mathcal{R}^T\end{array}\right)\left(\begin{array}{cc}1-\frac{1}{2}\zeta^\dagger\zeta&-\zeta^\dagger\\
\zeta&1-\frac{1}{2}\zeta\zeta^\dagger\end{array}\right).
\end{eqnarray}

Here, $\mathcal{S}$ and $\mathcal{R}$ are matrices used to diagonalize $M_\nu^{seesaw}$ and $\sqrt{2}{v}_{\eta}Y_X$.
\begin{eqnarray}
&&\hspace{0cm}\mathcal{S}^TM_\nu^{seesaw}\mathcal{S}=diag(m_{\nu1},m_{\nu2},m_{\nu3}),
\nonumber\\&&\hspace{0cm}\mathcal{R}^T\sqrt{2}{v}_{\eta}Y_X\mathcal{R}=diag(m_{\nu4},m_{\nu5},m_{\nu6}).
\end{eqnarray}
Here to make the discussion simple, we  assume that $Y_X$ is the diagonal matrix and $\mathcal{R} = 1$.
$M_\nu^{seesaw}$ is expressed as
\begin{eqnarray}
M_\nu^{seesaw}=\frac{v_u^2}{v_{\eta}}(Y_\nu^T)_{3\times3}(Y_X)_{3\times3}^{-1}(Y_\nu)_{3\times3}.
\end{eqnarray}

The neutrino mass mixing matrix is the sum of the tree-level result and the one-loop correction. It is expressed as follows(I,~J=1,~2,~3)
\begin{eqnarray}
M^{sum}=M_\nu+\delta m^{loop}_{ij}=&&\left(\begin{array}{cc}m^{loop}_{I,J}&\frac{1}{\sqrt{2}}{v}_{u}Y_{\nu}+m^{loop}_{I,J+3}
\\ \frac{1}{\sqrt{2}}v_{u}Y_{\nu}+m^{loop}_{I+3,J}&\sqrt{2}{v}_{\eta}Y_X+m^{loop}_{I+3,J+3}\end{array}\right).
\end{eqnarray}

Clearly, the matrix $M^{sum}$ includes one-loop correction that also has a seesaw structure.
At one loop level, we obtain a corrected effective light neutrino mass matrix of the form\cite{22sm10}
\begin{eqnarray}
&&\hspace{0cm}\mathcal{M}_\nu^{eff}\approx m^{loop}_{IJ}-(\frac{1}{\sqrt{2}}{v}_{u}Y_{\nu}+m^{loop}_{I,J+3})^T(\sqrt{2}{v}_{\eta}Y_{\nu}+m^{loop}_{I+3,J+3})^{-1}
\nonumber\\&&\hspace{2cm}\ \times(\frac{1}{\sqrt{2}}v_{u}Y_{\nu}+m^{loop}_{I+3,J}).
\end{eqnarray}

By the 'top-down' approach\cite{14yyl33,23sm18}, starting from the one-loop corrected light neutrino effective mass matrix $\mathcal{M}_\nu^{eff}$, one can obtain the corresponding Hermitian matrix.
\begin{eqnarray}
\mathcal{H}=(\mathcal{M}_\nu^{eff})^\dagger\mathcal{M}_\nu^{eff}.
\end{eqnarray}

By diagonalizing the $3\times3$ matrix $\mathcal{H}$, we can obtain three eigenvalues
\begin{eqnarray}
&&\hspace{0cm}m_1^2=\frac{a}{3}-\frac{1}{3}p(\cos\phi+\sqrt{3}\sin\phi),
\nonumber\\&&\hspace{0cm}m_2^2=\frac{a}{3}-\frac{1}{3}p(\cos\phi-\sqrt{3}\sin\phi),
\nonumber\\&&\hspace{0cm}m_3^2=\frac{a}{3}+\frac{2}{3}p\cos\phi.
\label{eee}
\end{eqnarray}

The specific form of the parameters in Eq.(\ref{eee}) is as follows
\begin{eqnarray}
&&\hspace{0cm}p=\sqrt{a^2-3b},~~~\phi=\frac{1}{3}\arccos\Big(\frac{1}{p^3}(a^3-\frac{9}{2}ab+\frac{27}{2}c)\Big),
~~~a=\mathrm{Tr}(\mathcal{H}),
\nonumber\\&&\hspace{0cm}b=\mathcal{H}_{11}\mathcal{H}_{22}
+\mathcal{H}_{11}\mathcal{H}_{33}+\mathcal{H}_{22}\mathcal{H}_{33}-\mathcal{H}^2_{12}-\mathcal{H}^2_{13}-\mathcal{H}^2_{23},~~~c=\mathrm{Det}(\mathcal{H}).
\end{eqnarray}

For the neutrino mass spectrum, there are two possibilities in the 3-neutrino mixing case. The neutrino mass spectrum with normal ordering (NO) is
\begin{eqnarray}
&&\hspace{-1cm}m_{\nu1}<m_{\nu2}<m_{\nu3},~~~m_{\nu1}^2=m_1^2,    ~~~m_{\nu2}^2=m_2^2,~~~m_{\nu3}^2=m_3^2,
\nonumber\\&&\hspace{-1cm}\Delta m^2_\odot=m_{\nu2}^2-m_{\nu1}^2=\frac{2}{\sqrt{3}}p\sin\phi>0,
\nonumber\\&&\hspace{-1cm}\Delta m^2_A=m_{\nu3}^2-m_{\nu2}^2=p(\cos\phi+{\frac{1}{\sqrt{3}}}\sin\phi)>0.
\end{eqnarray}

The neutrino  mass spectrum with inverse order (IO) is as follows
\begin{eqnarray}
&&\hspace{-1cm}m_{\nu3}<m_{\nu1}<m_{\nu2},~~~m_{\nu3}^2=m_1^2,    ~~~m_{\nu1}^2=m_2^2,~~~m_{\nu2}^2=m_3^2,
\nonumber\\&&\hspace{-1cm}\Delta m^2_\odot=m_{\nu2}^2-m_{\nu1}^2=p(\cos\phi-{\frac{1}{\sqrt{3}}}\sin\phi)>0,
\nonumber\\&&\hspace{-1cm}\Delta m^2_A=m_{\nu3}^2-m_{\nu2}^2=-p(\cos\phi+{\frac{1}{\sqrt{3}}}\sin\phi)<0.
\end{eqnarray}

The specific forms of the three mixing angles are shown in the Appendix \ref{C1}.

\section{Numerical analysis}
In this section, we discuss the numerical results of neutrino masses and mixing angles. We use the parameter space of the N-B-LSSM  and focus on the small neutrino Yukawa couplings. These couplings can contribute to the masses of light neutrinos at the tree level through the first type seesaw mechanism. The neutrino Yukawa coupling matrix $Y_\nu$ is a key parameter. It affects the masses of light neutrinos. Since the masses of light neutrinos are very small, we need to carefully consider and precisely calculate it. Therefore, the parameters used must be highly accurate. When performing the calculations, the following experimental limitations need to be taken into account:

1. We consider that the experimental constraint from the lightest CP-even Higgs mass is $125.20\pm0.11$ {\rm GeV}\cite{24s9,25s33,26s34}.

2. The $Z^\prime$ boson mass is larger than 5.1 TeV. The ratio between $M_{Z^\prime}$ and its gauge $M_{Z^\prime}/g_B \geq 6 ~{\rm TeV}$\cite{27s42}.

3. The new angle $\beta_\eta$ is constrained by LHC as $\tan\beta_\eta<1.5$.

4. The limitations for the particle masses are the following\cite{28g43,29g44,30g45,31g46,32g47,33g48}. The neutralino mass is limited to more than 116 GeV, the chargino mass is limited to more than 1100 GeV and the slepton mass is greater than 700 GeV.
Considering these limitations, we adopt the following parameters
\begin{eqnarray}
&&~~\tan{\beta}=15,~\tan{\beta}_{\eta}=0.62,~g_{B}=0.27,~g_{YB}=-0.18,
\nonumber\\&&~~\lambda_2=-0.25,~\lambda=0.3,~M_1=800~{\rm GeV},~M_2=1100~{\rm GeV},
\nonumber\\&&~~v_S=5~{\rm TeV},~\kappa=0.04,~M_{BL}=1000~{\rm GeV},~M_{BB'}=1100~{\rm GeV},
\nonumber\\&&~~v_\eta=37.6*\cos\theta_{\eta}~{\rm TeV},~~v_{\overline{\eta}}=37.6*\sin\theta_{\eta}~{\rm TeV},~~
\tan\theta_{\eta}=0.62,
\nonumber\\&&~~m_{\widetilde{L}ii}^2=1073^2~{\rm GeV^2},~m_{\widetilde{\nu}ii}^2=1\times10^6~{\rm GeV^2},~T_{Xii}=-4000~{\rm GeV},
\nonumber\\&&~~T_{\nu ii}=1000~{\rm GeV},~~T_{e ii}=1000~{\rm GeV},~~Y_{X ii}=0.43,
\nonumber\\&&~~m_{\widetilde{E}ii}^2=1.7\times10^6~{\rm GeV^2}~~(i=1,2,3).
\end{eqnarray}

In this study, the numerical results imply that
  the one-loop corrections contribute approximately 10\% to the total results including tree and one-loop contributions.
  Here we show the current experimental uncertainty and the future sensitivity in the table
  \ref{I}\cite{exp1, Jiangmen, DaYa},
  where we can find the current experimental uncertainties of
  neutrino mass squared difference and mixing angles vary from $1.1\%$ to $4.2\%$.
  The corresponding future sensitivities are proposed in the approximated region $[0.2\%,~ 3.4\%]$.
  In the end, one can find that the one-loop corrections are about
  $3\sim 5$ times as the current experiment sensitivities.
  Therefore, it is important to consider the one loop corrections to the neutrino mass and mixing.

\begin{table}
\caption{ Current experimental uncertainties and future sensitivities\cite{exp1, Jiangmen, DaYa}.}
\begin{tabular}{c|c|c|c|c}
\hline
parameters & Ordering &  PDG2024 & $1\sigma(\%)$ &Prospect(\%,~years)\\
\hline
$\frac{\Delta m_{21}^2}{10^{-5}{\rm eV}^2}$ & NO,~IO & $7.53\pm0.18$ &2.4 &(0.3,~6)\\
\hline
$\frac{|\Delta m_{32}^2|}{10^{-3}{\rm eV}^2}$ & NO & $2.455\pm0.028$&1.1 &(0.2,~6) \\

 & IO & $2.529\pm0.029$ &1.1 & \\
\hline
$\sin^2\theta_{12}$ & NO,~IO & $0.307\pm0.013$ &4.2 &(0.5,~6)\\
\hline
$\frac{\sin^2\theta_{13}}{10^{-2}}$ & NO,~IO & $2.19\pm0.07$ & 3.2 &(2.9,~-)\\
\hline
$\sin^2\theta_{23}$ & NO & $0.558^{+0.015}_{-0.024}$ &3.6 &(0.7-3.4,~10)\\
 & IO & $0.553^{+0.016}_{-0.024}$ &3.7 &\\
\hline
\end{tabular}
\label{I}
\end{table}

 The main effect is due to the tree-level seesaw contribution.
 We want to obtain an approximate analytic understanding of the result at tree level.
It would be good to explain the numerical results using analytic estimates.
So, we perform  approximate analytic analysis at normal order and inverse order.
\subsection{Approximate analysis of the mass matrix at tree level (NO)}
At first, we discuss the approximate analytic understanding for the NO condition.
Without any setting and approximation, the analytic results will be very complicated and impossible to read.
Therefore, we use the following relations to simplify the discussion
\begin{eqnarray}
(Y_{\nu})^{11}=-1.8\times(Y_{\nu})^{22},~~(Y_{\nu})^{23}=2.6\times(Y_{\nu})^{22},
~~(Y_{\nu})^{12}=(Y_{\nu})^{13}=1.5\times(Y_{\nu})^{33}.
\end{eqnarray}
Then the remaining two variables of the Yukawa coupling elements $(Y_\nu)^{ij}$ with $i,~j=1,~2,~3$ are
 $(Y_{\nu})^{22}$ and $(Y_{\nu})^{33}$. With the values for $v_u,~v_\eta,~Y_X$ in Eq.(37), when we adjust the values of the both variables $(Y_{\nu})^{22}$ and $(Y_{\nu})^{33}$,
 $\phi$ is a small value, stable around 0.0267. So, we simplify the formulas for neutrino mass in Eqs.(33)(35) with the assumption $\sin\phi\sim\phi\sim0.0267$ and $\cos\phi\sim1$,
\begin{eqnarray}
&&\hspace{0cm}m_1^2=\frac{a}{3}-\frac{1}{3}p(1+\sqrt{3}*0.0267),
\nonumber\\&&\hspace{0cm}m_2^2=\frac{a}{3}-\frac{1}{3}p(1-\sqrt{3}*0.0267),
\nonumber\\&&\hspace{0cm}m_3^2=\frac{a}{3}+\frac{2}{3}p,
 \nonumber\\
&&\Delta m_{\odot}^2 = m_{\nu_2}^2-m_{\nu_1}^2 ={2\over \sqrt{3}}p*0.0267,\nonumber\\
&&\Delta m_{A}^2 =m_{\nu_3}^2-m_{\nu_1}^2 =p(1+{1\over\sqrt{3}}*0.0267).
\end{eqnarray}
In this condition, $a$ and $p$ are
\begin{eqnarray}
&&a=\frac{1}{2v_\eta^2Y_X^2}(-129.933 \epsilon _{22}^4-27.04
   \epsilon _{22}^3 \epsilon_{33}-162.4 \epsilon _{22}^2
   \epsilon _{33}^2-30.6 \epsilon_{22} \epsilon _{33}^3-50.5
   \epsilon_{33}^4),
\nonumber\\
&&p=\frac{1}{2v_\eta^2Y_X^2}\Big(6856.37 \epsilon_{22}^8+9882.19
   \epsilon_{22}^7\epsilon_{33}+22477.8 \epsilon_{22}^6
   \epsilon_{33}^2+20979.9
   \epsilon_{22}^5 \epsilon_{33}^3\nonumber\\&&+22631.7 \epsilon_{22}^4
   \epsilon_{33}^4+13669.1
   \epsilon_{22}^3 \epsilon_{33}^5+8704.7 \epsilon_{22}^2
   \epsilon_{33}^6+2811.15
   \epsilon_{22} \epsilon_{33}^7+1016.31\epsilon_{33}^8\Big)^{1/2},
\end{eqnarray}
with $\epsilon_{ij}=\frac{v_u}{\sqrt{2}}(Y_\nu)^{ij}$  and $Y_{X11}\simeq Y_{X22}\simeq Y_{X33}\simeq Y_X$.

As $(Y_{\nu})^{22}=1.4\times10^{-6}$ and $(Y_{\nu})^{33}=5.0\times10^{-7}$, the approximate numerical results are
\begin{eqnarray}
&&m_{\nu_1}\simeq 1.743\times10^{-2} ~{\rm eV},~~~m_{\nu_2}\simeq 1.997\times10^{-2}~ {\rm eV},~~~
m_{\nu_3}\simeq 5.863\times10^{-2} ~{\rm eV},
 \nonumber\\
&&\Delta m_{\odot}^2 = 9.513\times 10^{-5}~ {\rm eV}^2,~~~\Delta m_{A}^2 =3.133\times 10^{-3}~ {\rm eV}^2.
\end{eqnarray}

\subsection{Approximate analysis of the mass matrix at tree level (IO)}
In the similar way, we study the tree level neutrino mass in the IO condition with following relations
\begin{eqnarray}
&&(Y_\nu)^{12}=-2.0\times(Y_\nu)^{22},~~~~(Y_\nu)^{11}= 1.4\times(Y_\nu)^{22}, \nonumber\\&&
(Y_\nu)^{33} = -0.2\times(Y_\nu)^{23},~~~~ (Y_\nu)^{13} = 1.5\times(Y_\nu)^{23}.
\end{eqnarray}
The remaining variables are $(Y_\nu)^{22}$ and $(Y_\nu)^{23}$, with the simplified results of $a$ and $p$
\begin{eqnarray}
&&a=\frac{1}{2v_\eta^2Y_X^2}(-106.602 \epsilon_{22}^4-16.04 \epsilon_{22}^2
   \epsilon_{23}^2-1.48 \epsilon_{22}
   \epsilon_{23}^3-21.6466 \epsilon_{23}^4),
\\
&&p=\frac{1}{2v_\eta^2Y_X^2}\Big(11226.8 \epsilon_{22}^8+855.576 \epsilon_{22}^6 \epsilon_{23}^2-314.412 \epsilon_{22}^5
   \epsilon_{23}^3-1578.84\epsilon_{22}^4
   \epsilon_{23}^4\nonumber\\&&\hspace{0.7cm}-63.6909 \epsilon_{22}^3
   \epsilon_{23}^5+314.746 \epsilon_{22}^2
   \epsilon_{23}^6+110.971 \epsilon_{22}
   \epsilon_{23}^7+133.876 \epsilon_{23}^8\Big)^{1/2}.
\end{eqnarray}
In this condition, $\phi$ is very near $\frac{\pi}{3}$ with the remaining two variables $(Y_\nu)^{22}$ and $(Y_\nu)^{23}$.
To deal with $\Delta m_{\odot}^2 $ better, we adopt $\phi=\frac{\pi}{3}-\delta\phi$,
and obtain the following approximate results keeping the first order of $\delta\phi$.
\begin{eqnarray}
&&m_1^2={a\over3}-{2\over3}p,
\nonumber\\
&&m_2^2={a\over3}-{1\over3}p(-1 + \sqrt{3} \delta\phi),
\nonumber\\
&&m_3^2={a\over3}+{2\over3}p(\frac{1}{2}+\frac{\sqrt{3}}{2}\delta\phi),\nonumber\\
&&\Delta m_{\odot}^2 = m_{\nu_2}^2-m_{\nu_1}^2 =p\frac{2\delta\phi}{\sqrt{3}}, \nonumber\\
&&\Delta m_{A}^2 = m_{\nu_3}^2-m_{\nu_2}^2 =-p(1+{1\over\sqrt{3}}\delta\phi).
\end{eqnarray}
Here, we use $(Y_\nu)^{22}=1.4\times10^{-6}$
and $(Y_\nu)^{23}=2.302\times10^{-6}$, then $\delta\phi$ is around 0.024.
The numerical results read as
\begin{eqnarray}
&&m_{\nu_1}\simeq 6.139\times10^{-2}~ {\rm eV},~~~m_{\nu_2}\simeq 6.147\times10^{-2} ~{\rm eV},~~~
m_{\nu_3}\simeq 2.169\times10^{-2} ~{\rm eV},
 \nonumber\\
&&\Delta m_{\odot}^2 = 9.154\times 10^{-5}~ {\rm eV}^2,~~~\Delta m_{A}^2 =3.348\times 10^{-3}~ {\rm eV}^2.
\end{eqnarray}

\subsection{NO spectrum numerical analysis}
To simplify the calculations, we have assumed that the Yukawa coupling matrix is a symmetric matrix $(Y_\nu=Y_\nu^T)$. This assumption reduces the number of independent parameters from 9 to 6. We further assume that $Y_\nu$ is a real matrix. The both assumptions do not affect the calculation substantively, which simplifies the computation.

Firstly, we study the neutrino mass spectrum at normal mass order (NO) by adjusting the model parameters. When the neutrino Yukawa coupling parameters are
\begin{eqnarray}
&&\hspace{0cm}(Y_{\nu})^{11}=-2.5143\times10^{-6},~~~~~~(Y_{\nu})^{22}=1.4\times10^{-6},
\nonumber\\&&\hspace{0cm}(Y_{\nu})^{33}=5.1348\times10^{-7},~~~~~~~~(Y_{\nu})^{12}=7.6533\times10^{-7},
\nonumber\\&&\hspace{0cm}(Y_{\nu})^{13}=7.9732\times10^{-7},~~~~~~~~(Y_{\nu})^{23}=3.6832\times10^{-6}.
\end{eqnarray}
The following values for the neutrino mass variances and mixing angles are obtained
\begin{eqnarray}
&&\hspace{0cm}\Delta m^2_A=2.47\times10^{-3}~{\rm eV^2},~~~\Delta m^2_\odot=7.89\times10^{-5}~{\rm eV^2},
\nonumber\\&&\hspace{0cm}m_{\nu_1} = 1.52 \times 10^{-2}~ {\rm eV},
~~m_{\nu_2} = 1.76 \times 10^{-2} ~{\rm eV},~~m_{\nu_3} = 5.20 \times 10^{-2}~ {\rm eV},
\nonumber\\&&\hspace{0cm}\sin^2\theta_{13}=0.0225,~~~\sin^2\theta_{12}=0.268,
~~~\sin^2\theta_{23}=0.562.
\end{eqnarray}

We scan pure loop-level parameters such as $g_{YB}$,
whose influence to the one-loop corrections is approximately at the 10\% level.
One-loop corrections account for approximately 10\% of the tree level results.
Therefore, in the summary, the imapct of $g_{YB}$ is estimateed to be around 1\% of the total results.
Due to this relatively minor variation,
it's hard to see the change of the result in the graph.
So, near the $3\sigma$ region of neutrino experiment data,
we choose the other parameters as variables to plot figures in the following.

\begin{figure}[h]
\setlength{\unitlength}{5mm}
\centering
\includegraphics[width=3.2in]{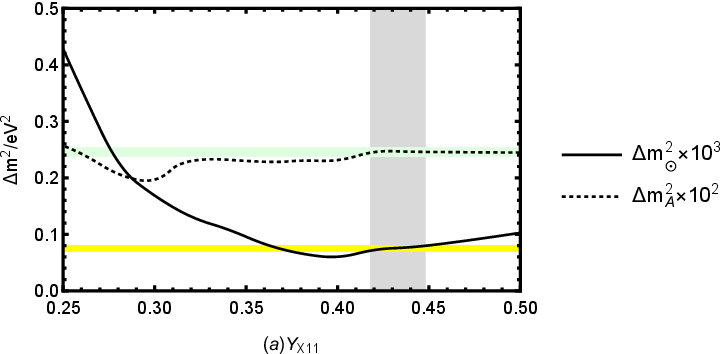}
\setlength{\unitlength}{5mm}
\centering
\includegraphics[width=3.1in]{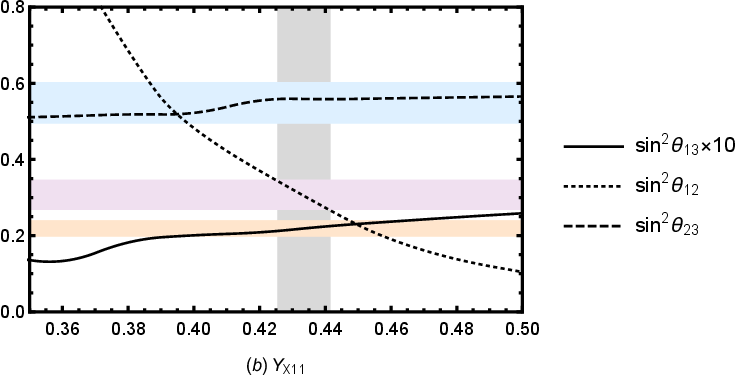}
\setlength{\unitlength}{5mm}
\caption{The neutrino mass-squared differences and mixing angles
versus $Y_{X11}$. The coloured bands in the figure represent experimental limits in the range of $3\sigma$.}{\label {T1}}
\end{figure}

$Y_X$ directly generates the right-handed neutrino mass term and indirectly modulates loop corrections to the neutrino mass matrix. Fig.\ref{T1} examines the impact of $Y_{X11}$ on neutrino mass-squared differences and mixing angles. Fig.\ref{T1}(a) displays the evolution of $\Delta m_A^2$ and $\Delta m_\odot^2$ across $Y_{X11} \in [0.25, 0.5]$. The mass-squared difference $\Delta m_A^2$ tends to change steadily after slowly decrease. In contrast, $\Delta m_\odot^2$ exhibits a clear decline trend then tends to be stable. Both $\Delta m_\odot^2$ and $\Delta m_A^2$ converge within experimental constraints (gray band) for $Y_{X11} \in [0.419, 0.45]$. Fig.\ref{T1}(b) displays the $Y_{X11}$ effects ($0.35$ to $0.5$) on mixing angles $\sin^2\theta_{12}$, $\sin^2\theta_{13}$, and $\sin^2\theta_{23}$. Both $\sin^2\theta_{13}$ and $\sin^2\theta_{23}$ demonstrate a gradual increasing trend. Conversely, $\sin^2\theta_{12}$ decreases clearly with increasing $Y_{X11}$. Crucially, all mixing angles simultaneously satisfy experimental limits within $Y_{X11} \in [0.425, 0.442]$ (gray region).
To ensure consistency with other  parameters, we take $Y_{X11} = 0.43$ as the benchmark value.

\begin{figure}[h]
\setlength{\unitlength}{5mm}
\centering
\includegraphics[width=3.26in]{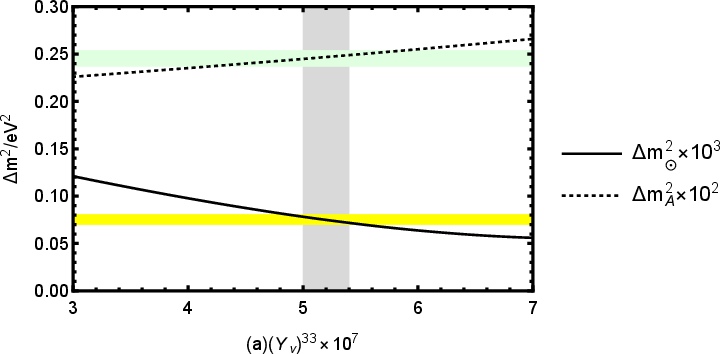}
\setlength{\unitlength}{5mm}
\centering
\includegraphics[width=3.05in]{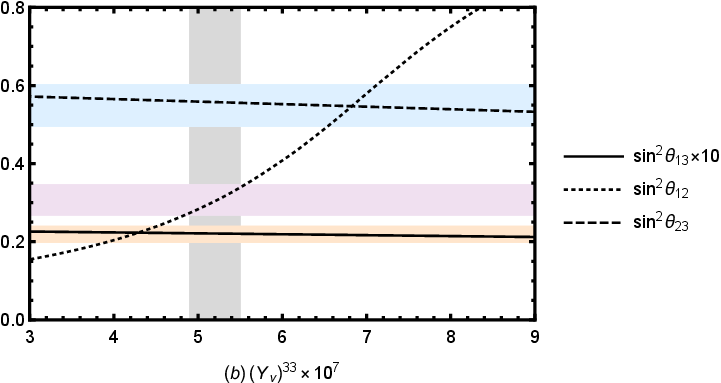}
\setlength{\unitlength}{5mm}
\caption{The neutrino mass-squared differences and mixing angles
versus $({Y_\nu})^{33}$. The coloured bands in the figure represent experimental limits in the range of $3\sigma$.}{\label {T2}}
\end{figure}

Despite the fact that neutrino Yukawa couplings coefficient are very small, they have a very significant effect on the neutrino mass and mixing angle, mainly due to their contribution at the tree level. Fig. \ref{T2}(a) shows the evolution of the  mass-squared differences ($\Delta m_A^2$ and $\Delta m_\odot^2$) across a scan of $(Y_{\nu})^{33}$ from $3\times10^{-7}$ to $7\times10^{-7}$. Over this domain, $\Delta m_A^2$ undergoes a slight rise. Conversely, $\Delta m_\odot^2$ shows a significant decrease. Crucially, both fall within experimentally allowed ranges (gray band) for $(Y_{\nu})^{33}$ values between $5\times10^{-7}$ and $5.4\times10^{-7}$. The corresponding variations in the neutrino mixing angles are plotted in Fig.\ref{T2}(b).  $\sin^2\theta_{13}$ decreases with increasing $(Y_{\nu})^{33}$, though the magnitude of change is marginal. $\sin^2\theta_{23}$ is also decreasing but the variation is slightly larger than that of $\sin^2\theta_{13}$. In contrast, $\sin^2\theta_{12}$ experiences a consistent upward change. At the benchmark point $(Y_{\nu})^{33} = 5.1348\times10^{-7}$, all neutrino mass-squared differences and mixing angles concurrently align with established experimental limits.
\begin{figure}[h]
\setlength{\unitlength}{5mm}
\centering
\includegraphics[width=3.3in]{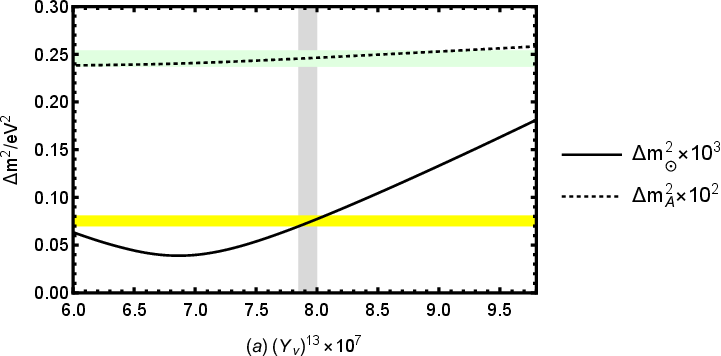}
\setlength{\unitlength}{5mm}
\centering
\includegraphics[width=3.05in]{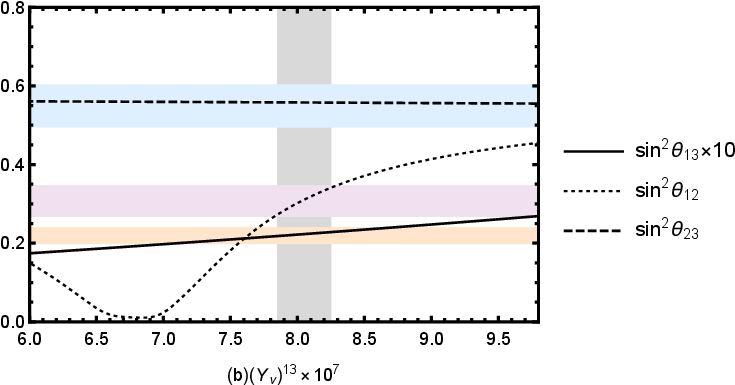}
\setlength{\unitlength}{5mm}
\caption{The neutrino mass-squared differences and mixing angles
versus $({Y_\nu})^{13}$. The coloured bands in the figure represent experimental limits in the range of $3\sigma$.}{\label {T3}}
\end{figure}

In addition to the diagonal factors, the non-diagonal elements of ${Y_\nu}$ are also very important in our discussion.  We investigate the impact of $(Y_{\nu})^{13}$ on theoretical predictions for neutrino mass-squared differences and mixing angles in Fig.\ref{T3}. Fig.\ref{T3}(a) displays the evolution of $\Delta m_A^2$ and $\Delta m_\odot^2$ with varying $(Y_{\nu})^{13}$.
The mass-squared difference $\Delta m_A^2$ exhibits a stable, near-linear rise across the parameter range. Conversely, $\Delta m_\odot^2$ demonstrates a two-phase evolution: first showing a slight decrease followed by a significant upward trend. Both quantities converge within experimental bounds (gray band) in the vicinity of $(Y_{\nu})^{13}=7.9732\times10^{-7}$. The corresponding mixing angle variations are shown in Fig.\ref{T3}(b). $\sin^2\theta_{13}$ exhibits a gradual increase and $\sin^2\theta_{23}$ exhibits fairly stable over the scanned range.
In contrast, $\sin^2\theta_{12}$ undergoes a different characteristic:
decreasing initially before trending upward at around $(Y_{\nu})^{13}=6.8\times 10^{-7}$. Remarkably, at the benchmark value $(Y_{\nu})^{13}=7.9732\times10^{-7}$, the theoretical predictions for mass-squared
differences and mixing angles simultaneously satisfy the experimental constraints.

\subsection{IO spectrum numerical analysis}
If the neutrino mass spectrum is IO, we use the following Yukawa coupling parameters
\begin{eqnarray}
&&\hspace{0cm}(Y_\nu)^{11}=1.89623\times10^{-6},~~~~~~(Y_\nu)^{22}=1.4\times10^{-6},
\nonumber\\&&\hspace{0cm}(Y_\nu)^{33}=-4.38536\times10^{-7},~~~~~~(Y_\nu)^{12}=-2.87584\times10^{-6},
\nonumber\\&&\hspace{0cm}(Y_\nu)^{13}=3.42168\times10^{-6},~~~~~~(Y_\nu)^{23}=2.30264\times10^{-6}.
\end{eqnarray}
Numerical results are obtained for the neutrino mass variance and mixing angle. The numerical results are as follows
\begin{eqnarray}
&&\hspace{0cm}\mid\Delta m^2_A\mid=2.53\times10^{-3}~{\rm eV^2},~~~\Delta m^2_\odot=7.605\times10^{-5}~{\rm eV^2},
\nonumber\\&&\hspace{0cm}m_{\nu_1} = 5.29 \times 10^{-2}~{\rm eV},~~~m_{\nu_2} = 5.36 \times 10^{-2}~{\rm eV},~~~m_{\nu_3} = 1.86 \times 10^{-2}~{\rm eV},
\nonumber\\&&\hspace{0cm}\sin^2\theta_{13}=0.0219,~~~\sin^2\theta_{12}=0.311,
~~~\sin^2\theta_{23}=0.552.
\end{eqnarray}

\begin{figure}[h]
\setlength{\unitlength}{5mm}
\centering
\includegraphics[width=3.25in]{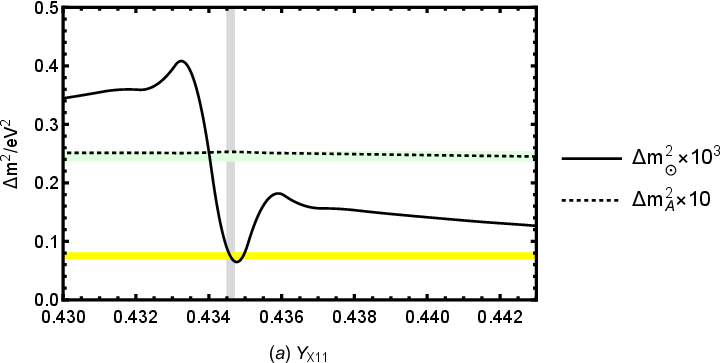}
\setlength{\unitlength}{5mm}
\centering
\includegraphics[width=3.1in]{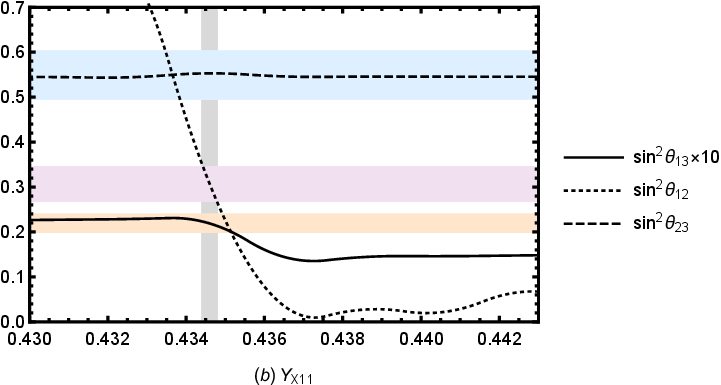}
\setlength{\unitlength}{5mm}
\caption{The neutrino mass-squared differences and mixing angles
versus $Y_{X11}$. The coloured bands in the figure represent experimental limits in the range of $3\sigma$.}{\label {T12}}
\end{figure}

Fig.\ref{T12}(a) exhibits the effect of $Y_{X11}$ on the mass variances $\Delta m_A^2$ and $\Delta m_\odot^2$, and the right panel shows the effect of $Y_{X11}$ on the three mixing angles $\sin^2\theta_{12}$, $\sin^2\theta_{13}$, and $\sin^2\theta_{23}$. In Fig.\ref{T12}(b), the trend of $\Delta m_\odot^2$ is shown to be relatively stable, and small changes in $Y_{X11}$ in the range $0.43-0.443$ have a significant effect on $\Delta m_A^2$. In Fig.\ref{T12}(b), it is shown that the value of $\sin^2\theta_{23}$ remains stable in the $Y_{X11}$ range $0.43-0.443$. $\sin^2\theta_{13}$ overall decreases slightly and then tends to stabilise. $\sin^2\theta_{12}$ decreases rapidly and then increases slowly in a wave-like manner. All three mixing angles near $Y_{X11}$=0.434 conform to the experimental limits.

\begin{figure}[h]
\setlength{\unitlength}{5mm}
\centering
\includegraphics[width=3.26in]{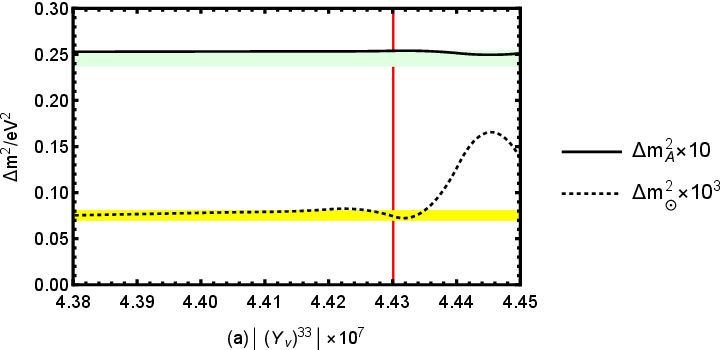}
\setlength{\unitlength}{5mm}
\centering
\includegraphics[width=3.05in]{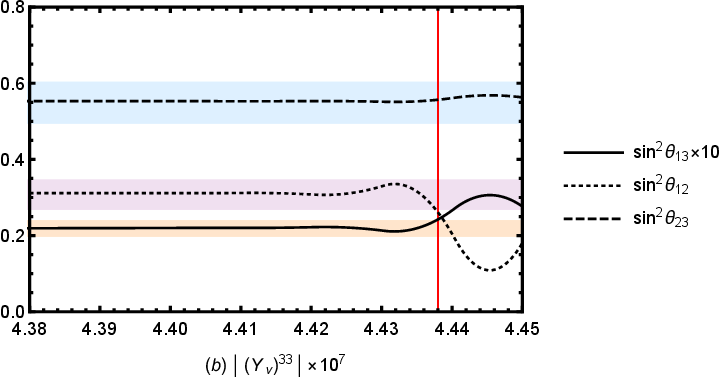}
\setlength{\unitlength}{5mm}
\caption{The neutrino mass-squared differences and mixing angles
versus $({Y_\nu})^{33}$. The coloured bands in the figure represent experimental limits in the range of $3\sigma$.}{\label {T13}}
\end{figure}

Here we discuss the effect of the diagonal elements of $Y_\nu$. Fig.\ref{T13} shows the effect of $({Y_\nu})^{33}$ on the neutrino mass variances and the three mixing angles. When $({Y_\nu})^{33}$ is in the range of $-4.45\times10^{-7}--4.38\times10^{-7}$, as a whole, the five values change relatively smoothly. When $({Y_\nu})^{33}$ is in the range of $-4.45\times10^{-7}--4.38\times10^{-7}$ all five values are in accordance with the experimental limits (i.e. left side of the red line). On the right side of the red line in Fig.\ref{T13}(a), $\Delta m_A^2$ stablely decreases and $\Delta m_\odot^2$ shows an increasing trend. On the right side of the red line in Fig.\ref{T13}(b), $\sin^2\theta_{23}$ changes more smoothly, $\sin^2\theta_{12}$ decreases and then increases, and $\sin^2\theta_{13}$ vice versa.

\begin{figure}[h]
\setlength{\unitlength}{5mm}
\centering
\includegraphics[width=3.3in]{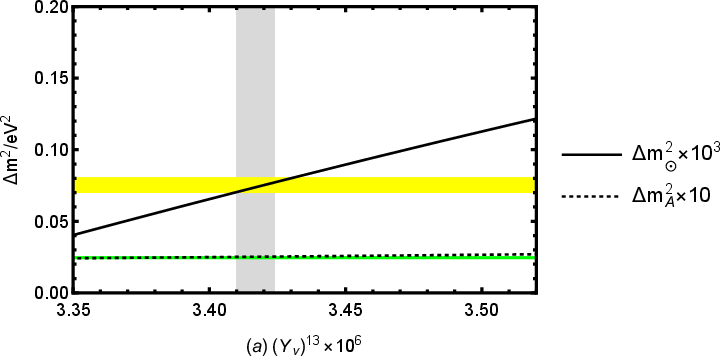}
\setlength{\unitlength}{5mm}
\centering
\includegraphics[width=3.05in]{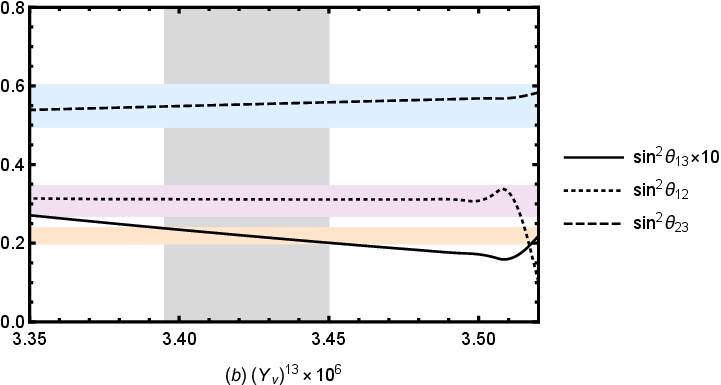}
\setlength{\unitlength}{5mm}
\caption{The neutrino mass squared differences and mixing angles
versus $({Y_\nu})^{13}$. The coloured bands in the figure represent experimental limits in the range of $3\sigma$.}{\label {T14}}
\end{figure}

We also discuss how the non-diagonal element $({Y_\nu})^{13}$ affects the theoretical predictions of the neutrino mixing angle and the mass-squared difference in Fig.\ref{T14}. As can be seen in the Fig.\ref{T14}(a), the change in $\Delta m_\odot^2$ is smoother and $({Y_\nu})^{13}$ has a significant effect on $\Delta m_A^2$. Both reach the experimental limit near $3.42\times10^{-6}$. The Fig.\ref{T14}(b) shows the effect of $({Y_\nu})^{13}$ on the three mixing angles. $\sin^2\theta_{13}$ shows a smooth decline followed by a rise. $\sin^2\theta_{23}$ rises slowly. $\sin^2\theta_{12}$ is more stable in the range of $3.35\times10^{-6}-3.50\times10^{-6}$, followed by a declining trend. To harmonise the other parameters, $({Y_\nu})^{13}$ takes the value of $3.42168\times10^{-6}$.

\section{Conclusion}
The SM cannot resolve the neutrino mass problem, and physicists believe that the SM should be a low-energy effective theory of large model. Therefore, the SM should be extended. The N-B-LSSM is a symmetry extension of the MSSM, in which we have investigated some processes in our previous work. The one-loop corrections to the neutrino mixing matrix are studied in this work.

In this paper, we systematically study the one-loop correction to the neutrino mass matrix in the N-B-LSSM and its fitting to experimental data on neutrino oscillations. We analyse the contributions to the neutrino mass from the loop diagrams of scalar lepton-chargino, scalar neutrino-neutralino and so on. Then we obtain the sum of tree-level and one-loop contributions to the neutrino mixing matrix. The one-loop corrected light neutrino effective mass matrix $M_{\nu}^{eff}$ is derived, and using the 'top-down' approach, we give formulas for the neutrino mass variances and mixing angles. For the neutrino mass spectra, NO and IO conditions are discussed numerically. The light neutrino masses are very tiny and there are five experiment constraints (three mixing angles and two mass squared differences), the obtained suitable parameter space is narrow. The numerical results show that the model can fit the current experimental data, including the neutrino mass squared deviation and the mixing angle, well within a specific parameter range.

 Our numercial results imply  that the one-loop corrections contribute
 approximately 10\% to the total result, and are 3 to 5 times larger than current experimental uncertainties. These demonstrate that one-loop corrections are becoming important for theory predictions for next-generation neutrino oscillation experiments which are expected to reach sub-percent precision for atmospheric and solar leptonic mixing parameters.

\appendix
\section{Mass matrix and coupling  in N-B-LSSM}\label{A1}
The mass matrix for chargino is
\begin{eqnarray}
m_{\chi^\pm} = \left(
\begin{array}{cc}
M_2&\frac{1}{\sqrt{2}}g_2v_\mu\\
\frac{1}{\sqrt{2}}g_2v_d&\frac{1}{\sqrt{2}}\lambda v_S\end{array}
\right).\label{Y2}
\end{eqnarray}

This matrix is diagonalized by U and V
\begin{eqnarray}
U^{*} m_{\chi^\pm} V^{\dagger}= m_{\chi^\pm}^{dia}.
 \end{eqnarray}

The mass matrix for slepton is
\begin{equation}
m^2_{\tilde{e}} = \left(
\begin{array}{cc}
m_{\tilde{e}_L\tilde{e}_L^*} &\frac{1}{\sqrt{2}}  v_d T_{e}^{\dagger}  -\frac{1}{2} v_u {\lambda} v_S Y_{e}^{\dagger} \\
\frac{1}{\sqrt{2}} v_d T_{e}  - \frac{1}{2}v_u {\lambda^*} v_S Y_{e}  &m_{\tilde{e}_R\tilde{e}_R^*}\end{array}
\right),
 \end{equation}
\begin{eqnarray}
&&m_{\tilde{e}_L\tilde{e}_L^*} = m_{\tilde{L}}^2+\frac{1}{8} \Big((g_{1}^{2} + g_{Y B}^{2}
+ g_{Y B} g_{B} -g_2^2)(v_{d}^{2}- v_{u}^{2})+ 2(g_{B}^2+ g_{Y B} g_{B})( v_{\eta}^{2}- v_{\bar{\eta}}^{2}
)
\Big)+\frac{v_{d}^{2}}{2} {Y_{e}^2} ,\nonumber\\&&
 %%%%%%%%%%%%%%%%%%%%%%%%%%%%%%%%%%%%%%%%%%%%%
m_{\tilde{e}_R\tilde{e}_R^*} = m_{\tilde{E}}^2-\frac{1}{8}  \Big([2(g_{1}^{2} + g_{Y B}^{2})+g_{Y B} g_{B}]
( v_{d}^{2}- v_{u}^{2})\nonumber\\&&\hspace{1.7cm}+(4g_{YB} g_{B}+2g_{B}^{2})(v_{\eta}^{2}- v_{\bar{\eta}}^{2})
\Big)+\frac{1}{2} v_{d}^{2} {  Y_{e}^2}.
\end{eqnarray}

The matrix is diagonalized by $Z^E$
\begin{eqnarray}
 Z^E m^2_{\tilde{e}} Z^{E,\dagger} = m^{diag}_{2,\tilde{e}}.
\end{eqnarray}

The mass matrix for CP-odd sneutrino is
\begin{equation}
m^2_{\nu^I}=\left(\begin{array}{cc}
m_{\sigma_L\sigma_L}&m_{\sigma_R\sigma_L}\\
m_{\sigma_L\sigma_R}&m_{\sigma_R\sigma_R}\end{array}
\right),
\end{equation}
\begin{eqnarray}
&&\hspace{0cm}m_{\sigma_L\sigma_L}=+\frac{1}{8}\mathbf{1}\Big(2g_B^2(-v_{\overline{\eta}}^2+v_\eta^2)+(g_1^2+g_{YB}^2+g_2^2)(-v_u^2+v_d^2)
\nonumber\\&&\hspace{0cm}+g_{YB}g_B(-2v_{\overline{\eta}}^2+2v_\eta^2-v_{\overline{\eta}}^2+v_\eta^2)\Big)+\frac{1}{4}
\Big(2v_u^2\mathfrak{R}(Y_\nu^TY_\nu^*)+4\mathfrak{R}(m_{\widetilde{L}}^2)\Big),
\nonumber\\&&\hspace{0cm}m_{\sigma_L\sigma_R}=\frac{1}{4}\Big(-2v_dv_S\mathfrak{R}(Y_\nu\lambda^*)+2v_u[-2v_\eta\mathfrak{R}(Y_X Y_\nu^*)+\sqrt{\mathfrak{R}(T_\nu)}]\Big),
\nonumber\\&&\hspace{0cm}m_{\sigma_R\sigma_R}=+\frac{1}{8}\mathbf{1}\Big(-2g_B^2(-v_{\overline{\eta}}^2+v_\eta^2)+g_{YB}g_B(-v_d^2+v_\eta^2)\Big)
\nonumber\\&&\hspace{0cm}+\frac{1}{4}\Big(2[2\mathfrak{R}(m_{\widetilde{\nu}}^2)
+2v_Sv_{\overline{\eta}}\mathfrak{R}(Y_X\lambda_2^*)]+2v_u^2\mathfrak{R}(Y_\nu Y_\nu^\dagger)-4\sqrt{2}v_\eta\mathfrak{R}(T_X)+8v_\eta^2\mathfrak{R}(Y_XY_X^*)\Big).
\end{eqnarray}

This matrix is diagonalized by $Z^I$
\begin{eqnarray}
 Z^I m^2_{\nu^I} Z^{I,\dagger} = m^{dia}_{2,{\nu^I}},
\end{eqnarray}

The mass matrix for CP-even sneutrino is
\begin{equation}
m^2_{\nu^R}=\left(\begin{array}{cc}
m_{\phi_L\phi_L}&m_{\phi_R\phi_L}^T\\
m_{\phi_L\phi_R}&m_{\phi_R\phi_R}\end{array}
\right),
\end{equation}
\begin{eqnarray}
&&\hspace{0cm}m_{\phi_L\phi_L}=+\frac{1}{8}\mathbf{1}\Big(2g_B^2(-v_{\overline{\eta}}^2+v_\eta^2)+(g_1^2+g_{YB}^2+g_2^2)(-v_u^2+v_d^2)
\nonumber\\&&\hspace{0cm}+g_{YB}g_B(-2v_{\overline{\eta}}^2+2v_\eta^2-v_{\overline{\eta}}^2+v_\eta^2)\Big)+\frac{1}{4}\Big(2v_u^2\mathfrak{R}(Y_\nu^TY_\nu^*)+4\mathfrak{R}(m_{\widetilde{L}}^2)\Big),
\nonumber\\&&\hspace{0cm}m_{\phi_L\phi_R}=\frac{1}{4}\Big(-2v_dv_S\mathfrak{R}(Y_\nu\lambda^*)+2v_u[2v_\eta\mathfrak{R}(Y_X Y_\nu^*)+\sqrt{\mathfrak{R}(T_\nu)}]\Big),
\nonumber\\&&\hspace{0cm}m_{\phi_R\phi_R}=+\frac{1}{8}\mathbf{1}\Big(-2g_B^2(-v_{\overline{\eta}}^2+v_\eta^2)+g_{YB}g_B(-v_d^2+v_\eta^2)\Big)
\nonumber\\&&\hspace{0cm}+\frac{1}{4}\Big(2\nu_u^2\mathfrak{R}(Y_\nu Y_\nu^\dagger)+4\mathfrak{R}(m_{\widetilde{\nu}}^2)+4\mathfrak{R}(m_{\widetilde{\nu}}^2)
\nonumber\\&&\hspace{0cm}+4v_\eta[2v_\eta\mathfrak{R}(Y_X Y_X^*)+\sqrt{2}\mathfrak{R}(T_X)]-4v_Sv_{\overline{\eta}}\mathfrak{R}(Y_X\lambda_2^*)\Big).
\end{eqnarray}

This matrix is diagonalized by $Z^R$
\begin{eqnarray}
 Z^R m^2_{\nu^R} Z^{R,\dagger} = m^{dia}_{2,{\nu^R}}.
\end{eqnarray}

\section{The definition of $F_1,F_2,F_3,F_4$}\label{B1}
\begin{eqnarray}
&&\hspace{0cm}F_1(m_1,m_2)=\frac{1}{16\pi^2}\Big(1+\frac{1}{(m_1^2-m_2^2)}(m_1^2\log{\frac{4\pi\mu^2}{m_1^2}}-m_2^2\log{\frac{4\pi\mu^2}{m_2^2}})\Big),
\nonumber\\&&\hspace{0cm}F_2(m_1,m_2)=\frac{1}{64\pi^2}
\Big(-\frac{m_1^2-3m_2^2}{2m_1^2-2m_2^2}+\frac{(m_1^4-2m_1^2m_2^2)\log{m_1^2}}{(m_1^2-m_2^2)^2}
\nonumber\\&&\hspace{2.5cm}+\frac{m_2^4\log{m_2^2}}{(m_1^2-m_2^2)^2}+\log{4\pi\mu^2}\Big),
\nonumber\\&&\hspace{0cm}F_3(m_1,m_2)=
\frac{1}{32(m_1^2-m_2^2)^3\pi^2}\Big(m_1^4-m_2^4-2m_1^2m_2^2\log{\frac{m_1^2}{m_2^2}}\Big),
\nonumber\\&&\hspace{0cm}F_4(m_1,m_2)=\frac{1}{96(m_1^2-m_2^2)^4\pi^2}(m_1^6-6m_1^4m_2^2+3m_1^2m_2^4+2m_2^6
+6m_1^2m_2^4\log{\frac{m_1^2}{m_2^2}}).
\end{eqnarray}

\section{Specific forms of the three mixing angles}\label{C1}
The mass-squared matrix $\mathcal{H}$ can be obtained to get the normalised eigenvectors as follows
\begin{eqnarray}
&&\hspace{0cm}\left(\begin{array}{c}(U_\nu)_{11} \\ (U_\nu)_{21}\\ (U_\nu)_{31}\end{array}\right)\;=\frac{1}{\sqrt{|X_1|^2+|Y_1|^2+|Z_1|^2}}\left(\begin{array}{c}X_1 \\ Y_1\\ Z_1\end{array}\right)\;,
\nonumber\\&&\hspace{0cm}\left(\begin{array}{c}(U_\nu)_{12} \\ (U_\nu)_{22}\\ (U_\nu)_{32}\end{array}\right)\;=\frac{1}{\sqrt{|X_2|^2+|Y_2|^2+|Z_2|^2}}\left(\begin{array}{c}X_2 \\ Y_2\\ Z_2\end{array}\right)\;,
\nonumber\\&&\hspace{0cm}\left(\begin{array}{c}(U_\nu)_{13} \\ (U_\nu)_{23}\\ (U_\nu)_{33}\end{array}\right)\;=\frac{1}{\sqrt{|X_3|^2+|Y_3|^2+|Z_3|^2}}\left(\begin{array}{c}X_3 \\ Y_3\\ Z_3\end{array}\right)\;.
\end{eqnarray}

The specific forms of $X_I, Y_I, Z_I$ when $I = 1, 2, 3$ are as follows
\begin{eqnarray}
&&\hspace{0cm}X_1=(\mathcal{H}_{22}-m_{\nu1}^2)(\mathcal{H}_{33}-m_{\nu1}^2)-\mathcal{H}_{23}^2,
~~~Y_1=\mathcal{H}_{13}\mathcal{H}_{23}-\mathcal{H}_{12}(\mathcal{H}_{33}-m_{\nu1}^2),
\nonumber\\&&\hspace{0cm}Z_1=\mathcal{H}_{12}\mathcal{H}_{23}-\mathcal{H}_{13}(\mathcal{H}_{22}-m_{\nu1}^2),
~~~~~~~~~X_2=\mathcal{H}_{13}\mathcal{H}_{23}-\mathcal{H}_{12}(\mathcal{H}_{33}-m_{\nu1}^2),
\nonumber\\&&\hspace{0cm}Y_2=(\mathcal{H}_{11}-m_{\nu2}^2)(\mathcal{H}_{33}-m_{\nu2}^2)-\mathcal{H}_{13}^2,
~~~Z_2=\mathcal{H}_{12}\mathcal{H}_{13}-\mathcal{H}_{23}(\mathcal{H}_{11}-m_{\nu2}^2),
\nonumber\\&&\hspace{0cm}X_3=\mathcal{H}_{12}\mathcal{H}_{23}-\mathcal{H}_{13}(\mathcal{H}_{22}-m_{\nu3}^2),
~~~~~~~~~Y_3=\mathcal{H}_{12}\mathcal{H}_{13}-\mathcal{H}_{23}(\mathcal{H}_{11}-m_{\nu3}^2),
\nonumber\\&&\hspace{0cm}Z_3=(\mathcal{H}_{11}-m_{\nu3}^2)(\mathcal{H}_{22}-m_{\nu3}^2)-\mathcal{H}_{12}^2.
\end{eqnarray}

The mixing angles among three tiny neutrinos can be defined as follows
\begin{eqnarray}
&&\hspace{0cm}\sin\theta_{13}=|(U_\nu)_{13}|,
~~~~~~~~~~~~\cos\theta_{13}=\sqrt{1-|(U_\nu)_{13}|^2},
\nonumber\\&&\hspace{0cm}\sin\theta_{23}=\frac{|(U_\nu)_{23}|}{\sqrt{1-|(U_\nu)_{13}|^2}},
~~~\cos\theta_{23}=\frac{|(U_\nu)_{33}|}{\sqrt{1-|(U_\nu)_{13}|^2}},
\nonumber\\&&\hspace{0cm}\sin\theta_{12}=\frac{|(U_\nu)_{12}|}{\sqrt{1-|(U_\nu)_{13}|^2}},
~~~\cos\theta_{12}=\frac{|(U_\nu)_{11}|}{\sqrt{1-|(U_\nu)_{13}|^2}}.
\end{eqnarray}

\begin{acknowledgments}

This work is supported by National Natural Science Foundation of China (NNSFC)
(No.12075074), Natural Science Foundation of Hebei Province
(A2023201040, A2022201022, A2022201017, A2023201041), Natural Science Foundation of
Hebei Education Department (QN2022173), Post-graduate's Innovation
Fund Project of Hebei University (HBU2024SS042), the Project of the China
Scholarship Council (CSC) No. 202408130113. X. Dong acknowledges support from Funda\c{c}\~{a}o para a Ci\^{e}ncia e a Tecnologia (FCT, Portugal) through the projects CFTP FCT Unit UIDB/00777/2020 and UIDP/00777/2020.

\end{acknowledgments}


\begin{thebibliography}{50}

\vspace{3mm}
\bibitem{1sm2} T2K Collab, Phys. Rev. Lett. \textbf{107} (2011) 041801;
MINOS Collab, Phys. Rev. Lett. \textbf{107} (2011) 181802;
DOUBLE-CHOOZ Collab, Phys. Rev. Lett. \textbf{108} (2012) 131801;
DAYA-BAY Collab, Phys. Rev. Lett. \textbf{108} (2012) 171803;
PoS HQL \textbf{2014} (2014) 019.
\bibitem{2s5} I. Girardi , S.T. Petcov , A.V. Titov, Nucl. Phys. B \textbf{894} (2015) 733-768.
\bibitem{3s6} P. Ghosh, S. Roy, J. High Energy Phys. \textbf{0904} (2009) 069.
\bibitem{4s7} P. Ghosh, P. Dey, B. Mukhopadhyaya, S. Roy, J. High Energy Phys.  \textbf{1005} (2010) 087.
\bibitem{DayaBay1} J.~Li, PoS \textbf{EPS-HEP2023} (2024) 148
\bibitem{DayaBay2}D.~Adey, \textit{et al.} [DayaBay] Phys. Rev. Lett. \textbf{121} (2018)  241805.
\bibitem{DayaBay3}F.~P.~An, \textit{et al.} [DayaBay] Phys. Rev. Lett. \textbf{108} (2012) 171803.
\bibitem{lep1}V.~Cirigliano, K.~Fuyuto, C.~Lee, \textit{et al.} J. High Energy Phys. \textbf{03} (2021) 256.
\bibitem{6pdg} S.~Navas \textit{et al.} [Particle Data Group], Phys. Rev. D \textbf{110} (2024) no.3, 030001.
\bibitem{7h} X.Y.Han, S.M.Zhao, L. Ruan \textit{et al.} Eur. Phys. J. C \textbf{85} (2025) 2, 163.
\bibitem{8seesaw} A. Batra, P. Bharadwaj, S. Mandal \textit{et al.} J. High Energy Phys. \textbf{07} (2023) 221.
\bibitem{9sm7} N. Escudero, D.E.L. Fogliani, C. Munoz, \textit{et al.} J. High Energy Phys. \textbf{12} (2008) 099.
\bibitem{102zlxzqqb} E.~Ma, Phys. Rev. D \textbf{73} (2006) 077301.
\bibitem{11Minkowski}P.~Minkowski, Phys. Lett. B \textbf{67} (1977) 421-428.
\bibitem{10yyl22}P. Ghosh, P. Dey, B. Mukhopadhyaya \textit{et al.} J. High Energy Phys.  \textbf{05} (2010) 087;
P. Ghosh, S. Roy, J. High Energy Phys. \textbf{04} (2009) 069.
\bibitem{11sm} S.M. Zhao, T.F.~Feng, X.X.~Dong \textit{et al.} Nucl. Phys. B \textbf{910} (2016) 225-239.
\bibitem{12tff} T.F. Feng, X.Q. Li, Phys. Rev. D {\bf63} (2001) 073006.
\bibitem{13yyl} Y.L. Yan, T.F. Feng, J.L. Yang \textit{et al.} Phys. Rev. D \textbf{97} (2018) no.5, 055036.
\bibitem{101zlxz}
M.~Dvornikov,Phys. Rev. D \textbf{111} (2025) no.5, 056009.
\bibitem{14yyl33} H.B.~Zhang, T.F.~Feng, L.N.~Kou \textit{et al.} Int. J. Mod. Phys. A \textbf{28} (2013) no.24, 1350117.
\bibitem{15tff8} Y. Grossman, H.E. Haber, Phys. Rev. D  \textbf{59} (1999) 093008.
\bibitem{16s13} U.~Ellwanger, C.~Hugonie, A.M.~Teixeira, Phys. Rept. \textbf{496} (2010) 1-77.
\bibitem{17h32} G. Belanger, J.D. Silva, H.M. Tran, Phys. Rev. D \textbf{95} (2017) 115017.
\bibitem{18h33} V. Barger, P.F. Perez, S. Spinner, Phys. Rev. Lett. \textbf{102} (2009) 181802.
\bibitem{19h34} P.H. Chankowski, S. Pokorski, J. Wagner, Eur. Phys. J. C \textbf{47} (2006) 187.
\bibitem{20h35} J.L. Yang, T.F. Feng, S.M. Zhao, \textit{et al.} Eur. Phys. J. C \textbf{78} (2018) 714.
\bibitem{21tff12} J. Liu, Y.P. Yao, Phys. Rev. D \textbf{41} (1990) 2147;
H. Simma, D. Wyler, Nucl. Phys. B \textbf{344} (1990) 283;
 S. Herrlich and J. Kalinowski, Nucl. Phys. B \textbf{381} (1992) 50.
\bibitem{22sm10} P. Ghosh, P. Dey, B. Mukhopadhyaya, \textit{et al.} J. High Energy Phys. \textbf{05} (2010) 087.
\bibitem{23sm18} B. Dziewit, S. Zajac, M. Zralek, Acta Phys. Pol. B \textbf{42} (2011) 2509.
\bibitem{24s9}S. Navas et al., Phys. Rev. D \textbf{110} (2024) 3, 030001.
\bibitem{25s33}CMS collaboration, Phys. Lett. B \textbf{716} (2012) 30.
\bibitem{26s34}ATLAS collaboration, Phys. Lett. B \textbf{716} (2012) 1.
\bibitem{27s42}ATLAS collaboration, Phys. Lett. B \textbf{796} (2019) 68.
\bibitem{28g43}P. Cox, C.C. Han, T.T. Yanagida, Phys. Rev. D \textbf{104} (2021) 075035.
\bibitem{29g44} M. V. Beekveld, W. Beenakker, M. Schutten, \textit{et al.} SciPost Phys. \textbf{11} (2021) 3, 049.
\bibitem{30g45} M. Chakraborti, L. Roszkowski, S. Trojanowski, J. High Energy Phys. \textbf{05} (2021) 252.
\bibitem{31g46} F. Wang, L. Wu, Y. Xiao, \textit{et al.} Nucl. Phys. B \textbf{970} (2021) 115486.
\bibitem{32g47} M. Chakraborti, S. Heinemeyer, I. Saha, Eur. Phys. J. C \textbf{81} (2021) 12.
\bibitem{33g48} M. Endo, K. Hamaguchi, S. Iwamoto, \textit{et al.} J. High Energy Phys. \textbf{07} (2021) 075.
\bibitem{exp1}J.~Zhang, J.~Cao, JHEP \textbf{03} (2023) 072, [arXiv:2206.15317 [hep-ex]].
\bibitem{Jiangmen}A.~Abusleme, \textit{et al.} [JUNO], Chin. Phys. C \textbf{46} (2022)  123001,
[arXiv:2204.13249 [hep-ex]].
\bibitem{DaYa}K.B. Luk, Reactor neutrino i$_-$latest results from daya bay, June, 2022. 10.5281/zenodo.6683712.
\end{thebibliography}
\end{document}